\documentclass[useAMS,usenatbib]{mn2e}

\usepackage{color}
\usepackage{txfonts}
\usepackage{rotating}

\def\twelveCO{\mbox{$^{12}$CO}}
\def\thirteenCO{\mbox{$^{13}$CO}}
\def\CeighteenO{\mbox{C$^{18}$O}}

\def\HthirteenCOplus{\mbox{H$^{13}$CO$^+$}}
\def\HCOplus{\mbox{HCO$^+$}}
\def\Msun{M$_{\odot}$}
\def\cmcube{cm$^{-3}$}
\def\arcsecs{$^{\prime\prime}$}
\def\arcmin{$^{\prime}$}
\def\antemp{$T_{\rm A}^*$}
\def\kms{km~s$^{-1}$}
\def\clfind{{\sc clfind}}
\def\gclumps{{\sc gaussclumps}}
\def\sigmaoned{$\sigma_{1 \rm D}$}
\def\sigmactc{$\sigma_{\nu}$}
\def\ctcdisp{interclump dispersions}
\long\def\changed#1{\textcolor{black}{#1}}

\title[NGC~2068 --- Structure and Kinematics]{The structure and kinematics of dense gas in NGC~2068}
\author[S. L. Walker-Smith et. al.]{S. L. Walker-Smith$^{1}$\thanks{E-mail:
sw547@mrao.cam.ac.uk}, J. S. Richer$^{1,2}$, J. V. Buckle$^{1,2}$, R. J. Smith$^{3}$, J. S. Greaves$^{4}$ \newauthor and I. A. Bonnell$^{4}$\\
$^{1}$Astrophysics Group, Cavendish Laboratory, J J Thomson Avenue, Cambridge, CB3 0HE\\
$^{2}$Kavli Institute for Cosmology, Institute of Astronomy, University of Cambridge, Madingley Road, Cambridge, CB3 0HA\\
$^{3}$Zentrum f\"{u}r Astronomie der Universit\"{a}t Heidelberg, Institut f\"{u}r Theoretische Astrophysik, Albert-Ueberle-Str. 2, Heidelberg, Germany\\
$^{4}$School of Physics and Astronomy, St Andrews University, North Haugh, St Andrews, Fife, KY16 9SS}

\begin{document}

\date{}

\pagerange{\pageref{firstpage}--\pageref{lastpage}} \pubyear{2012}

\maketitle

\label{firstpage}

\begin{abstract}
We have carried out a survey of the NGC~2068 region in the Orion B molecular cloud using HARP on the JCMT, in the \thirteenCO\ and \CeighteenO\ ($J = $ 3 -- 2) and \HthirteenCOplus\ ($J = $ 4 -- 3) lines. We used \thirteenCO\ to map the outflows in the region, and matched them with previously defined SCUBA cores. We decomposed the \CeighteenO\ and \HthirteenCOplus\ into Gaussian clumps, finding 26 and 8 clumps respectively. The average deconvolved radii of these clumps is $6200 \pm 2000$~AU and $3600 \pm 900$~AU for \CeighteenO\ and \HthirteenCOplus\ respectively. We have also calculated virial and gas masses for these clumps, and hence determined how bound they are. We find that the \CeighteenO\ clumps are more bound than the \HthirteenCOplus\ clumps (average gas mass to virial mass ratio of 4.9 compared to 1.4). We measure clump internal velocity dispersions of $0.28 \pm 0.02$ \kms\ and $0.27 \pm 0.04$ \kms\ for \CeighteenO\ and \HthirteenCOplus\ respectively, although the \HthirteenCOplus\ values are heavily weighted by a majority of the clumps being protostellar, and hence having intrinsically greater linewidths. We suggest that the starless clumps correspond to local turbulence minima, and we find that our clumps are consistent with formation by gravoturbulent fragmentation. We also calculate inter-clump velocity dispersions of $0.39 \pm 0.05$ \kms\ and $0.28 \pm 0.08$ \kms\ for \CeighteenO\ and \HthirteenCOplus\ respectively. The velocity dispersions (both internal and external) for our clumps match results from numerical simulations of decaying turbulence in a molecular cloud. However, there is still insufficient evidence to conclusively determine the type of turbulence and timescale of star formation, due to the small size of our sample.
\end{abstract}


\section{Introduction}

An understanding of the properties of molecular gas is important in the investigation of star formation in our Galaxy. A complete theory of star formation should be able to explain certain basic properties and characteristics of dense cores and protostars. The kinematics of the dense cores and their surrounding natal environment, as well as the relation of the Core Mass Function (CMF) to the stellar Initial Mass Function (IMF), are all linked to the initial conditions of star formation. Most stars form in clusters \citep{2003ARA&A..41...57L}, although \citet{2010MNRAS.409L..54B} find that the percentage of stars found in clusters appears to be dependent on the definitions used. Many studies of the less dense gas ({e.g.} CO low-J transitions) in these clusters exist ({e.g.} \citet{1998ApJS..118..455O}, \citet{2002AAS...201.3406D}, \citet{2008ApJ...686..384D}). However, there are fewer surveys of the very dense gas tracers ({e.g.} HCN or \HCOplus\ $J =$ 4 -- 3 transitions), which reveal the density peaks in the gas most closely associated with the protostars. With the advent of receivers like HARP on the JCMT \citep{2009MNRAS.399.1026B}, \changed{and HERA on IRAM \citep{2004A&A...423.1171S}}, large-scale dense gas surveys are now feasible.

In recent years, a general consensus has been reached that star formation is a dynamical process, occurring within a few freefall times; the method by which molecular gas clouds collapse to form protostellar cores is also generally accepted to be a combination of gravity and turbulence. This \emph{gravoturbulent fragmentation} is essentially a two-phase process \citep{2011EAS....51..133K}: the turbulence produces density enhancements in the molecular gas cloud through a network of interacting shocks; these density enhancements decrease the Jeans mass locally \citep{2005MNRAS.361....2C} allowing the densest regions (formed at the shock intersection points) to collapse, accrete mass and form stars. \changed{It should be noted that although these turbulent shocks are expected to be widespread in molecular clouds, there is no observational evidence for them as yet. The overall picture is complicated by the presence of magnetic fields, which play a role on large scales, and cause modifications to the core collapse processes \citep{2008MNRAS.385.1820P}.}

Although turbulence plays such an important part in star formation, its origins, length- and timescales are still not understood. It is unclear whether turbulence in star formation regions decays freely in a crossing time \citep{2000ApJ...530..277E}, or is driven \citep{2001ApJ...556..837K} and therefore persists for a longer timescale. Most competitive accretion simulations \citep{2006MNRAS.370..488B,2009MNRAS.396..830S} feature molecular clouds initially supported by turbulence that decays on a timescale comparable to the crossing time of the molecular cloud. This allows competitive accretion to start on small scales and grow progressively larger, fed by loss of kinetic energy from the ambient cloud. \citet{2001ApJ...556..837K} however have shown evidence for competitive accretion in an environment with driven turbulence; as long as the turbulence is driven on large enough scales, clusters can form where gas and stars are virialised. They have also produced representative core mass functions from their numerical simulations; however they find that with these alone they are unable to distinguish between the driven and decaying modes of turbulence. Indeed, the general properties of the cores produced in differing turbulent environments at low resolution are insufficently different \citep{2009ApJ...704L.124O} to dismiss a particular environment. 

We have therefore decided to use an alternative method --- examining the \emph{kinematics} of the dense gas cores, rather than their spatial and structural properties. \citet{2008AJ....136..404O} have performed comparisons of the kinematic properties of protostellar and starless cores produced in simulations that utilise both driven and decaying turbulence modes. Comparing the intercore velocity dispersions ({i.e.} the dispersions of the core centroid velocities), they found that protostellar cores in decaying turbulence simulations have a higher intercore velocity dispersion than the ambient gas dispersion; protostellar cores and starless cores in driven turbulence simulations have subvirial intercore velocity dispersions and intercore velocity dispersions close to that of the gas respectively. They therefore suggest that comparing the starless and protostellar intercore velocity dispersions to the net gas dispersion could potentially distinguish between the two environments observationally. 

We have chosen to observe the \HthirteenCOplus\ $J=  $ 4 -- 3 transition as it is optically thin, traces the densest regions (n$_{\rm crit} \sim 10^6$ cm$^{-3}$) and will be most likely to trace the dense inner regions of prestellar cores. We can calculate linewidths and virial ratios of \HthirteenCOplus\ cores to determine how well bound they are; we can also potentially use the core-to-core velocity dispersions to distinguish between the two turbulence scenarios. We have also observed the \CeighteenO\ $J = $ 3 -- 2 transition which is optically thin over most of the region and will also trace more of the less dense core envelopes for comparison with the \HthirteenCOplus. In addition, we observed the \thirteenCO\ $J = $ 3 -- 2 transition to investigate the large-scale kinematics, \changed{as it should trace more of the bulk gas in the region than the other two molecules}, as well as providing evidence for outflows which would pinpoint protostellar cores.

\subsection{NGC~2068 region}
\begin{figure}
\centering
\includegraphics[width=8.3cm]{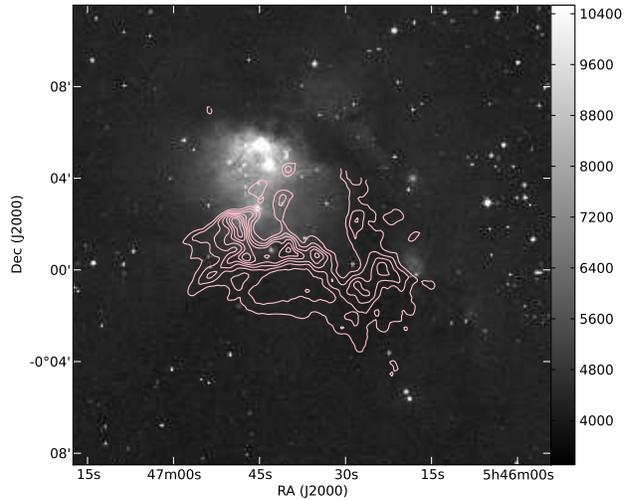}
\caption{A DSS2-IR near-infrared image of NGC~2068, with our \CeighteenO\ emission contours overlaid, integrated over velocity range 8.5 -- 12.5 \kms. Contours are from 2 -- 9 K~\kms, at 1 K~\kms\ intervals.}
\label{ir_overlay}
\end{figure}

The Orion B molecular cloud complex is the closest region of high-mass star formation, and lies at a distance of around 415 pc \citep{1982AJ.....87.1213A,2007A&A...474..515M}. It includes 5 major star formation regions, one of which, NGC~2068, is the subject of this paper. NGC~2068 is a bright reflection nebula containing an infrared cluster; submillimetre emission associated with the nebula has been observed to the south, as shown in Figure \ref{ir_overlay}. 

The first unbiased molecular gas survey of NGC~2068 was carried out by \citet{1991ApJ...368..432L}, hereafter LBS90, using CS $J = $ 2 -- 1 emission. They identified 2 clumps --- LBS 10 and LBS 17 --- at 1.7\arcmin\ resolution. The region has since been mapped by many others, both in molecular line emission ({e.g.} \citet{2009ApJ...691.1560I}, in \HthirteenCOplus\ $J= $ 1 -- 0) as well as in SCUBA 850 $\umu$m dust emission ({e.g.} \citet{2007MNRAS.374.1413N}, hereafter NWT), breaking the two LBS regions into smaller clumps and cores. \citet{2001ApJ...556..215M} identified 18 smaller clumps with diameters of $\sim$ 25\arcsecs\ using both 850 $\umu$m dust emission and CO isotopologue mapping. The clumps they identified are labelled OriBsmm (for Orion B submmillimetre emission) and then numbered by decreasing declination. \citet{2001A&A...372L..41M}, hereafter M01, identified 31 cores with FWHM diameters of $\sim$13\arcsecs\ from  850 $\umu$m dust emission; their cores are labelled LBSX-MMY, where X is the corresponding LBS clump that the core belongs to (either 10 or 17) and Y begins at 1 and increases with RA. 

The NGC~2068 region has been extensively studied and well-documented at different densities and resolutions, and is a fairly small, self-contained region with obvious dust and gas structure. We have therefore chosen this region for investigations of the kinematics of pre- and protostellar cores and their surrounding gas, in order to examine the initial conditions for star formation and determine the possible mechanisms for star formation in this region.

\subsection{Outline}
We present an analysis of spectral line data from NGC~2068, examining the gaseous structure of the region, as well as its kinematics. We break the region down into Gaussian clumps using two molecules that trace different densities, to examine the kinematics and mechanisms leading to the fragmentation and formation of prestellar cores. Section \ref{sec:obs} gives an overview of the observations and data reduction procedure. Section \ref{sec:res} describes the reduced data sets, highlighting the multiple-velocity-component structure of the gas, and several high-velocity outflows. Section \ref{sec:clfind} presents the clumpfinding analysis of the region, using \CeighteenO\ and \HthirteenCOplus, while Section \ref{sec:clprop} examines the characteristic properties of the clumps, including shape and size, masses and velocity dispersions. We also present a comparison of observational data with numerical simulations of star formation. We summarise the results and conclude in Section \ref{sec:concl}.

\section{Observations}\label{sec:obs}
\subsection{Description of observations}
The observations were made over a total of 12 nights between 6 September and 2 October 2010, using HARP at the JCMT \citep{2009MNRAS.399.1026B}. In total, 11.5 hours of data were taken, with an area of 400\arcsecs\ $\times$ 600\arcsecs\ (centred on postion $05^{\rm h}46^{\rm m}40^{\rm s}$, +00$^{\circ}$00\arcmin00\arcsecs \changed{(J2000)}) being mapped. 

All the standard telescope observing procedures were followed for each night of observations, with regular pointings and focusing of the JCMT's secondary mirror. Standard spectra were also taken towards various well-known calibration sources, to verify that the intensity in the tracking receptor matched recorded standards to within calibration tolerance before subsequent observations were allowed to continue. The fully-sampled maps were taken in raster position-switched mode, an `on-the-fly' data collection method where the HARP array continuously scans in a direction parallel to the sides of the map \changed{to produce a full-sampled map of the area} \citep{2009MNRAS.399.1026B}. The telescope is pointed at an off-position \changed{($05^{\rm h}43^{\rm m}44^{\rm s}$, +00$^{\circ}$21\arcmin42.2\arcsecs (J2000 coordinates))} after every row in the map to obtain background values that are subtracted from the raw data. A low-resolution \twelveCO\ $J =$ 3 -- 2 `stare' observation was first performed towards this position to verify that it was indeed emission-free. Data are presented in units of corrected antenna temperature \antemp, which is related to the main beam temperature ($T_{\rm mb}$) using $T_{\rm mb}$ = \antemp/$\eta_{\rm mb}$ \citep{2000..RW..book}. A value of $\eta_{\rm mb}$ = 0.61 was used, following \citet{2009MNRAS.399.1026B}. 

Emission from 3 molecules --- \thirteenCO, \CeighteenO\ and \HthirteenCOplus\ --- was observed and details of the transitions and frequencies are given in Table \ref{mol_trans}. The frequencies and energies quoted are taken from LAMBDA, the Leiden Atomic and Molecular Database \footnote{http://home.strw.leidenuniv.nl/~moldata/}. The \thirteenCO/\CeighteenO\ data were observed simultaneously, utilising the ACSIS high-resolution dual sub-band mode. Each sub-band was divided into 4096 61~kHz channels, giving a velocity resolution of 0.055~\kms. The \HthirteenCOplus\ data was observed in single sub-band mode, splitting the 250~MHz bandwidth into 8192 channels, giving a velocity resolution of 0.026~\kms. \changed{The bandwidth of the observations is around 200~\kms, and has a range of -90~\kms\ to 120~\kms. Overall, the weather was good. The mean system and receiver temperatures were 330.5~K and 90~K (at 345~GHz) respectively, and the opacity at 225~GHz ranged from 0.065 to 0.075.}

\begin{table}
\caption{Molecules observed in this particular run for NGC~2068, with their transitions (Column 2), frequencies (Column 3) and energies of the upper level above ground (Column 4). The critical densities of each transition (Column 5) are taken from \citet{2000..RW..book}.}
\begin{center}
\begin{tabular}{ccccc}
\hline
Molecule & $\Delta J$ & $\nu_{\rm trans}$/~GHz & $E_{\rm u}$/~K & $n_{\rm crit}$/~\cmcube \\ \hline
\thirteenCO\ & 3 -- 2 & 330.58 & 31.7 & 2 $\times$ 10$^4$\\
\CeighteenO\  & 3 -- 2 & 329.33 & 31.6 & 2 $\times$ 10$^4$\\ 
\HthirteenCOplus\ & 4 -- 3 & 346.99 & 41.6 & 3 $\times$ 10$^7$ \\ \hline
\end{tabular}
\end{center}
\label{mol_trans}
\end{table}

\begin{figure*}
\centerline{\includegraphics[width=8.0cm,height=6.5cm]{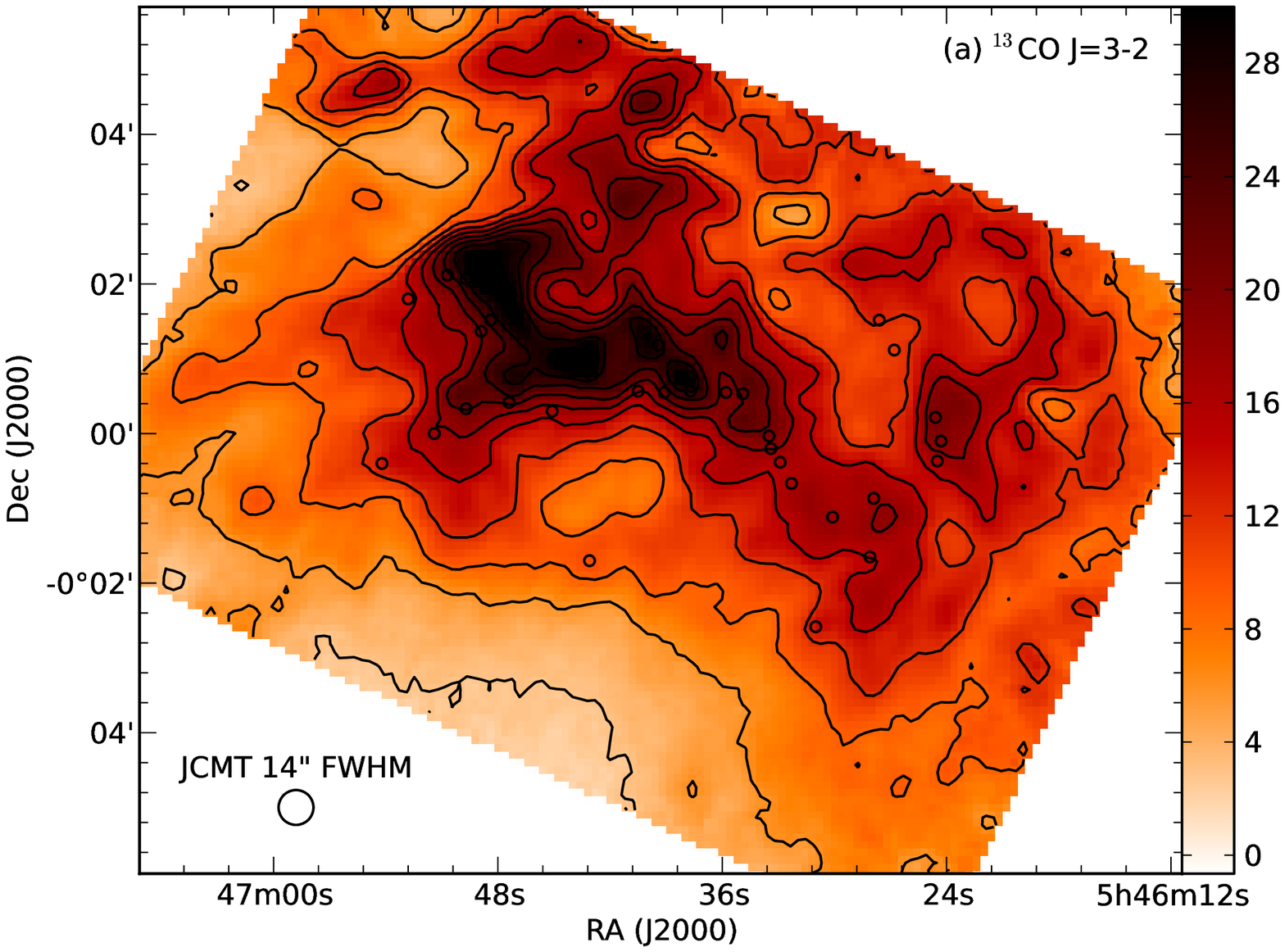}\qquad
	    \includegraphics[width=8.0cm,height=6.5cm]{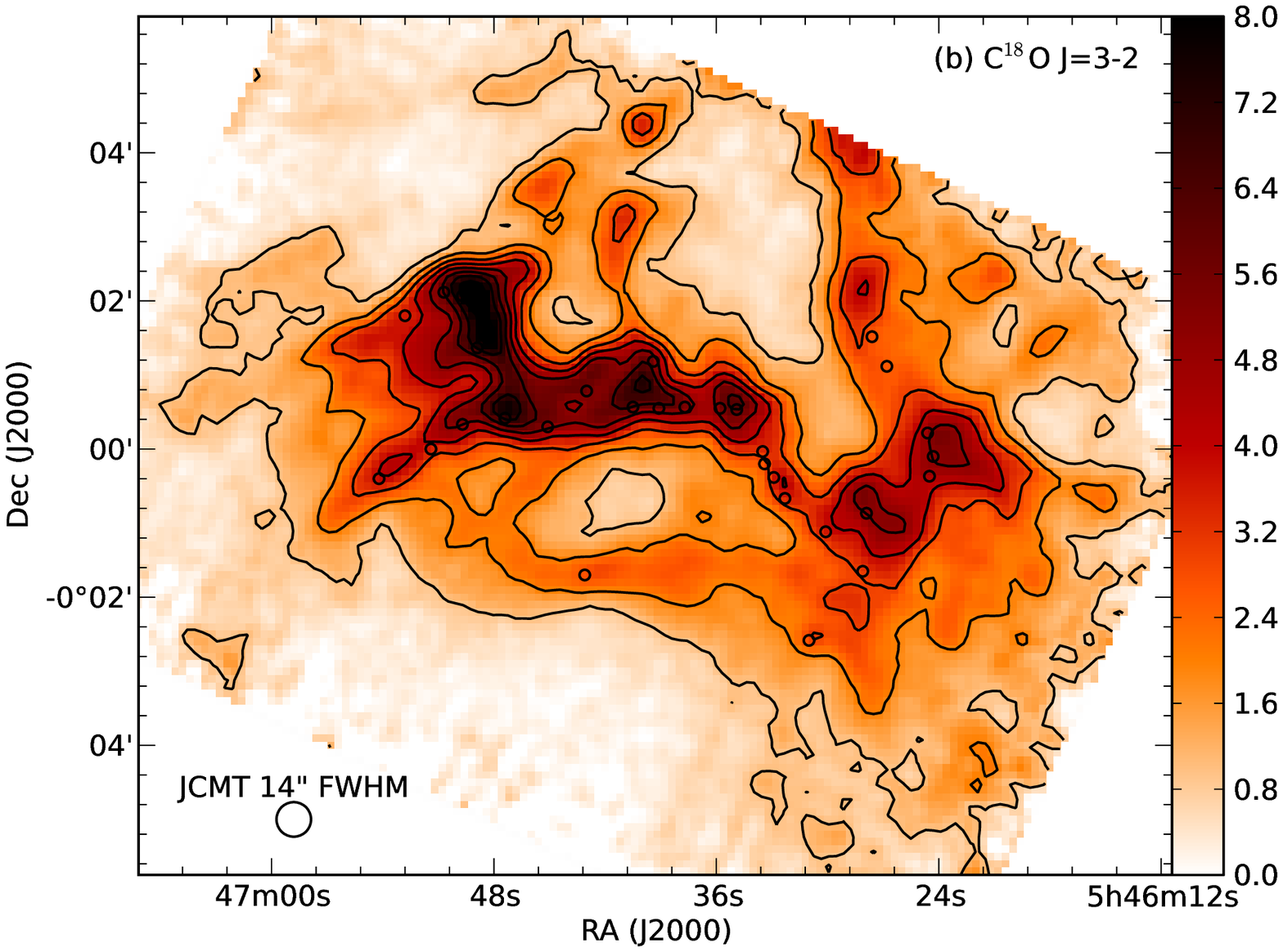}}
\smallskip
\centerline{\includegraphics[width=8.0cm,height=6.5cm]{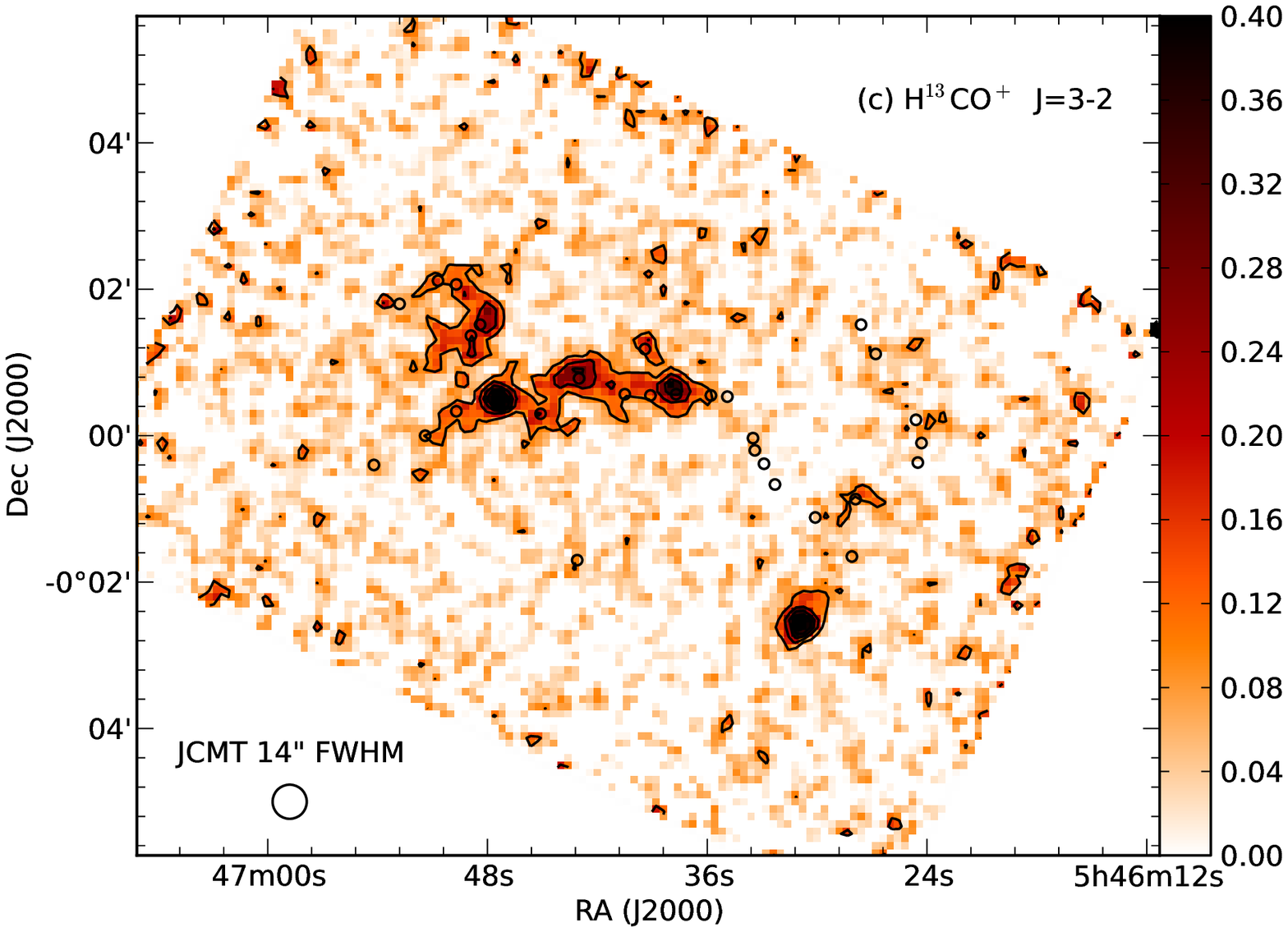}\qquad
    	    \includegraphics[width=8.0cm,height=6.5cm]{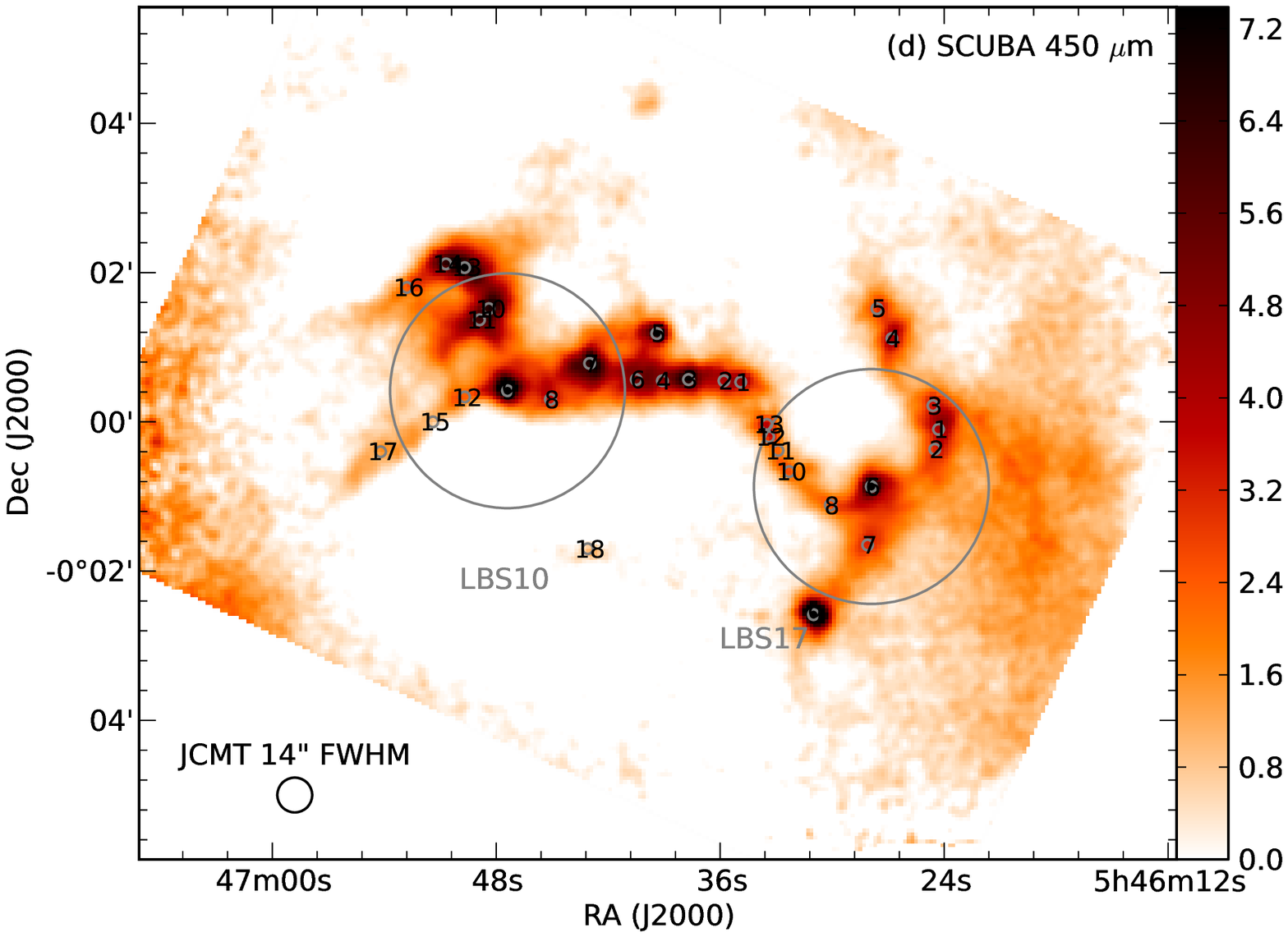}}
\caption{(a) Integrated Intensity of \thirteenCO\ (contours from 3 -- 30 K~\kms\ in 3 K~\kms\ intervals); (b) \CeighteenO\ (contours from 1 -- 9 K~\kms\ in 1 K~\kms\ intervals); (c) \HthirteenCOplus\ (contours from 0.1 -- 0.5 K~\kms\ in 0.1 K~\kms\ intervals). Crosses show positions of SCUBA cores identified by M01. (d) SCUBA 450 $\umu$m emission obtained from the CADC archives with M01 core positions, and clump positions for LBS 10 and 17, marked with circles and the corresponding clump/core number.}
\label{int_intensity}
\end{figure*}

\begin{figure*}
\centerline{\includegraphics[width=8.0cm,height=6.5cm]{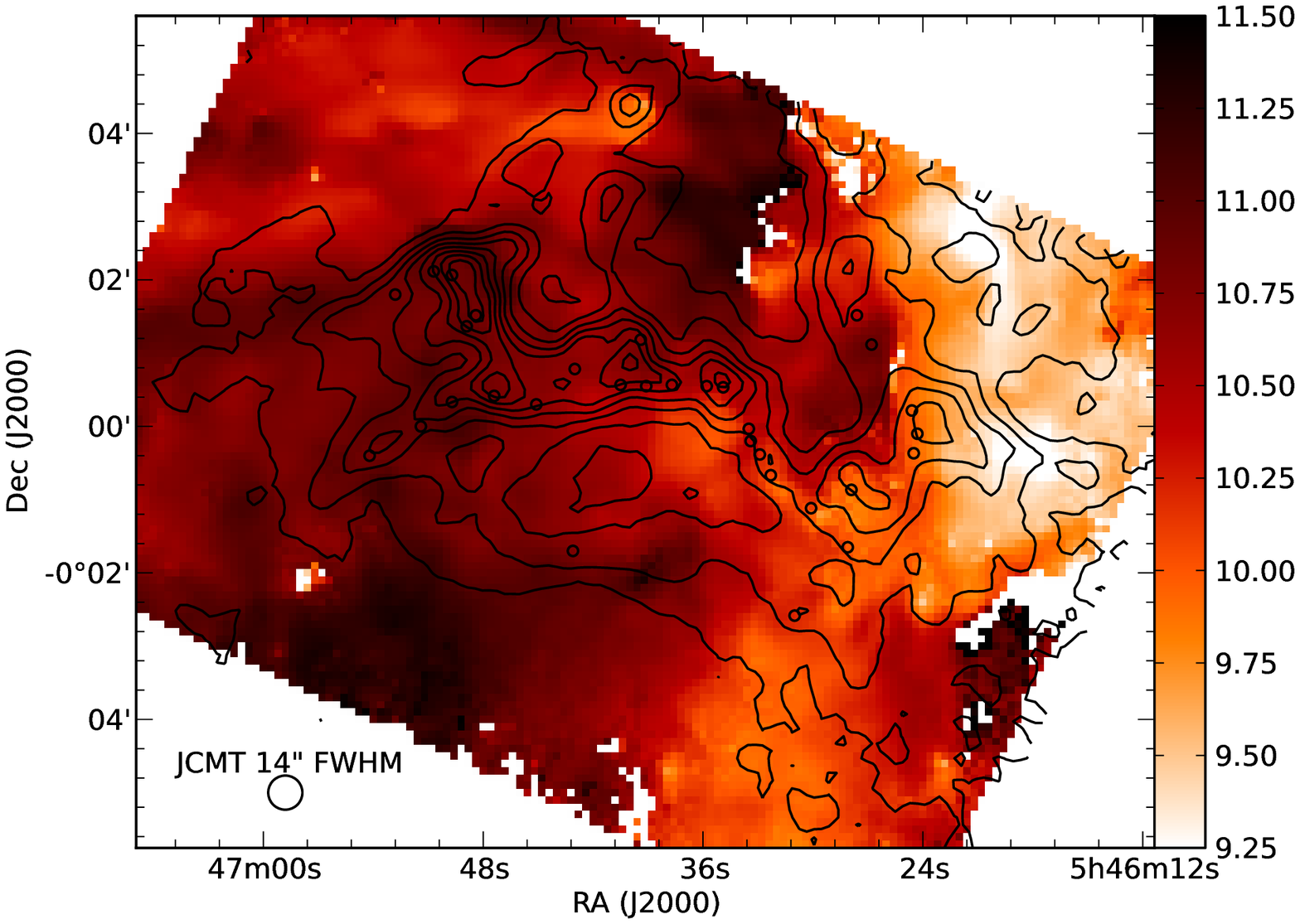}\qquad
	    \includegraphics[width=8.0cm,height=6.5cm]{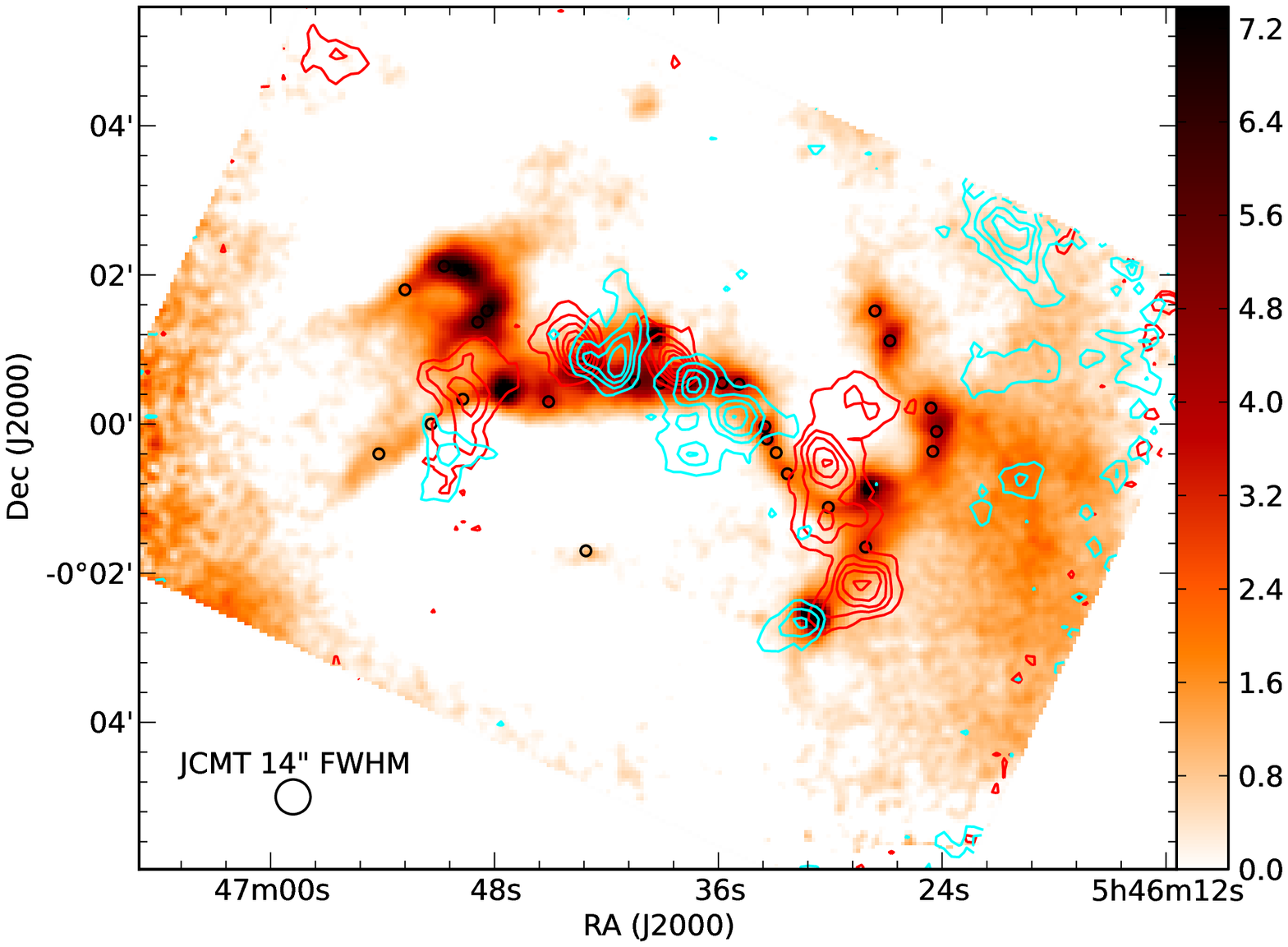}}
\caption{Left: A velocity map for \thirteenCO\ of the line centre velocity at each spatial pixel. Overlaid are \thirteenCO\ integrated intensity contours (integrated between 8.5 -- 12.5 \kms). Right: SCUBA 450 $\umu$m emission, with \thirteenCO\ red and blue line wing contours overlaid. Black circles indicate positions of the M01 SCUBA cores, and magenta diamonds show protostellar (Class I) cores as determined by NWT from Spitzer IRAC data.}
\label{kinematics}
\end{figure*}

\subsection{Data reduction}
The data were reduced using the Starlink project\footnote{http://starlink.jach.hawaii.edu/starlink} software, using {\sc smurf makecube} routines to convert from time-series to spectral cubes and {\sc kappa} routines to mask out poorly-performing detectors ({\sc chpix}), remove linear baselines ({\sc mfittrend}) and alter the image resolution ({\sc sqorst}). 

The final data were sampled onto a 6\arcsecs\ per pixel cube, using Gaussian gridding kernels with FWHM 9\arcsecs\ for all three molecules. Taking into account the JCMT beamwidth (14\arcsecs\ at 345~GHz \citep{2009MNRAS.399.1026B}), the effective FWHM for the cubes is 17.7\arcsecs\ for \thirteenCO\ and \CeighteenO, and 16.8\arcsecs\ for \HthirteenCOplus. The data were also rebinned to a velocity resolution of 0.1 \kms. \changed{At this resolution, the mean 1$\sigma$ rms noise on the maps for the three molecules was 0.10~K, 0.08~K and 0.03~K for \thirteenCO, \CeighteenO\ and \HthirteenCOplus\ respectively.}

\section{Results}\label{sec:res}

\subsection{Integrated intensity and structure}

Figure \ref{int_intensity} shows the integrated intensity $\int{T_A^*{\rm d}v}$ images of the three molecules (all integrated between 8.5 and 12.5 \kms). The SCUBA 450 $\umu$m emission obtained from the CADC Legacy Catalogue \citep{2008ApJS..175..277D} is also included for comparison. 

Of the three molecules, the \CeighteenO\ data most closely resembles the optically thin SCUBA dust emission. The large-scale filamentary structure in the region that is seen in the dust emission is also seen in the \CeighteenO: there is a clear central ridge running E--W which contains several smaller dense condensations, and a second large filament running almost perpendicular to the ridge in the west of the region. Some smaller filaments can also be seen, running diagonally off the main ridge to the southeast. A cavity structure centred on $05^{\rm h}46^{\rm m}40^{\rm s}$, $-00^{\circ}01^{\prime}$ can also be seen in the \CeighteenO, and to a lesser extent in the \thirteenCO. The \thirteenCO\ emission (being the most abundant) covers most of the region, but demonstrates little filamentary structure. The brightest emission originates in the centre of the region, corresponding to the LBS 10 clump. The \HthirteenCOplus\ emission, by contrast, is concentrated in small, dense clumps, with no extended emission. Almost all of the emission lies in the central region, although the brightest emission is seen to the west --- a feature which is conspicuously absent in the CO isotopologues. 

\subsection{Velocity structure and kinematics}

The left-hand panel of Figure \ref{kinematics} shows the line centre velocity (as determined by a best-fit Gaussian to the \thirteenCO\ spectrum at each spatial pixel) over the entire region. We can see that there is a large-scale velocity gradient in a W-E direction over the region --- emission in the west generally peaks between 9.0 to 9.5 \kms\ while emission in the east generally peaks between 10.5 to 11.0 \kms\ --- which could indicate differential rotation across the region as a whole.

\begin{figure}
\centerline{\includegraphics[width=3.4cm,angle=-90]{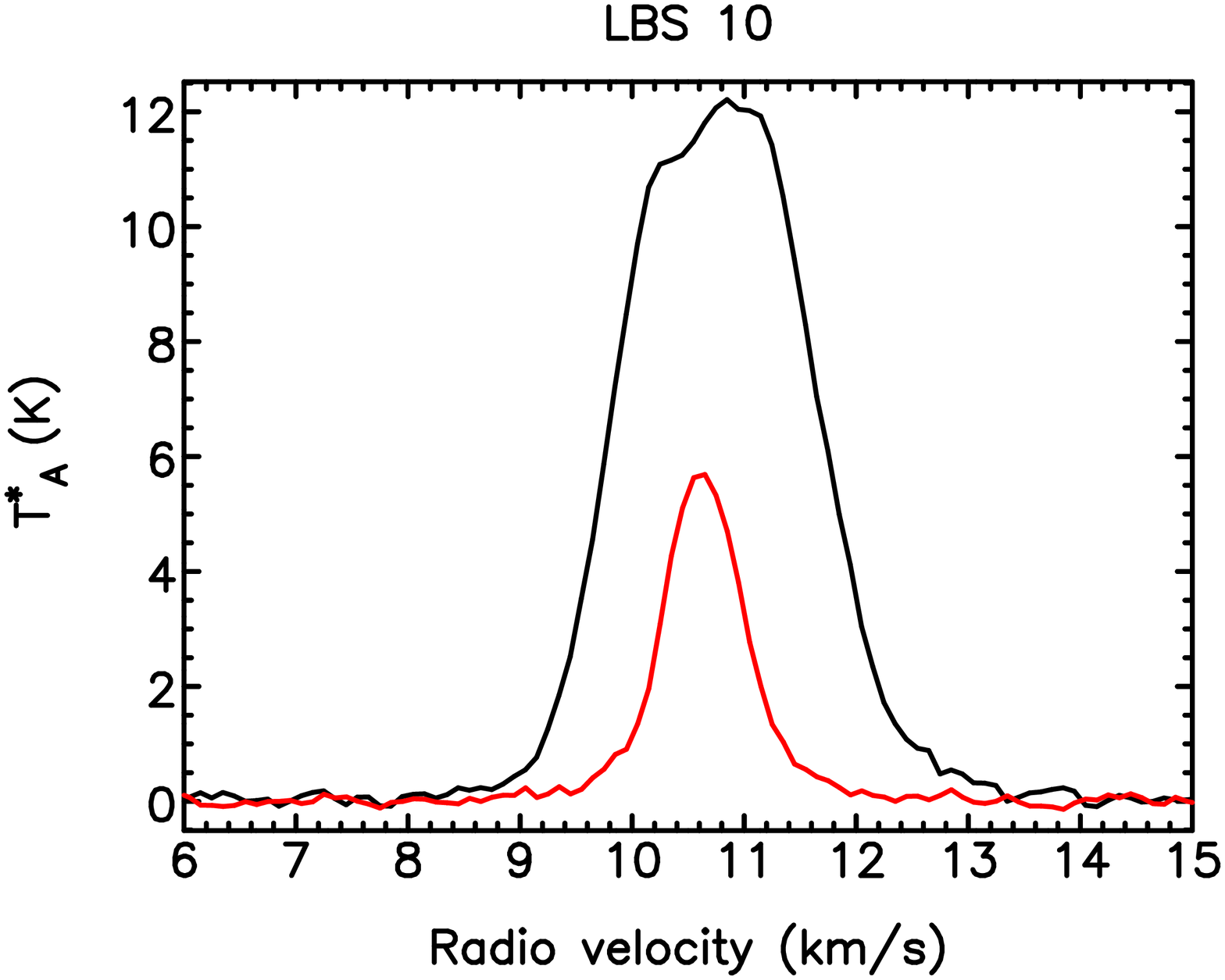}\quad
            \includegraphics[width=3.4cm,angle=-90]{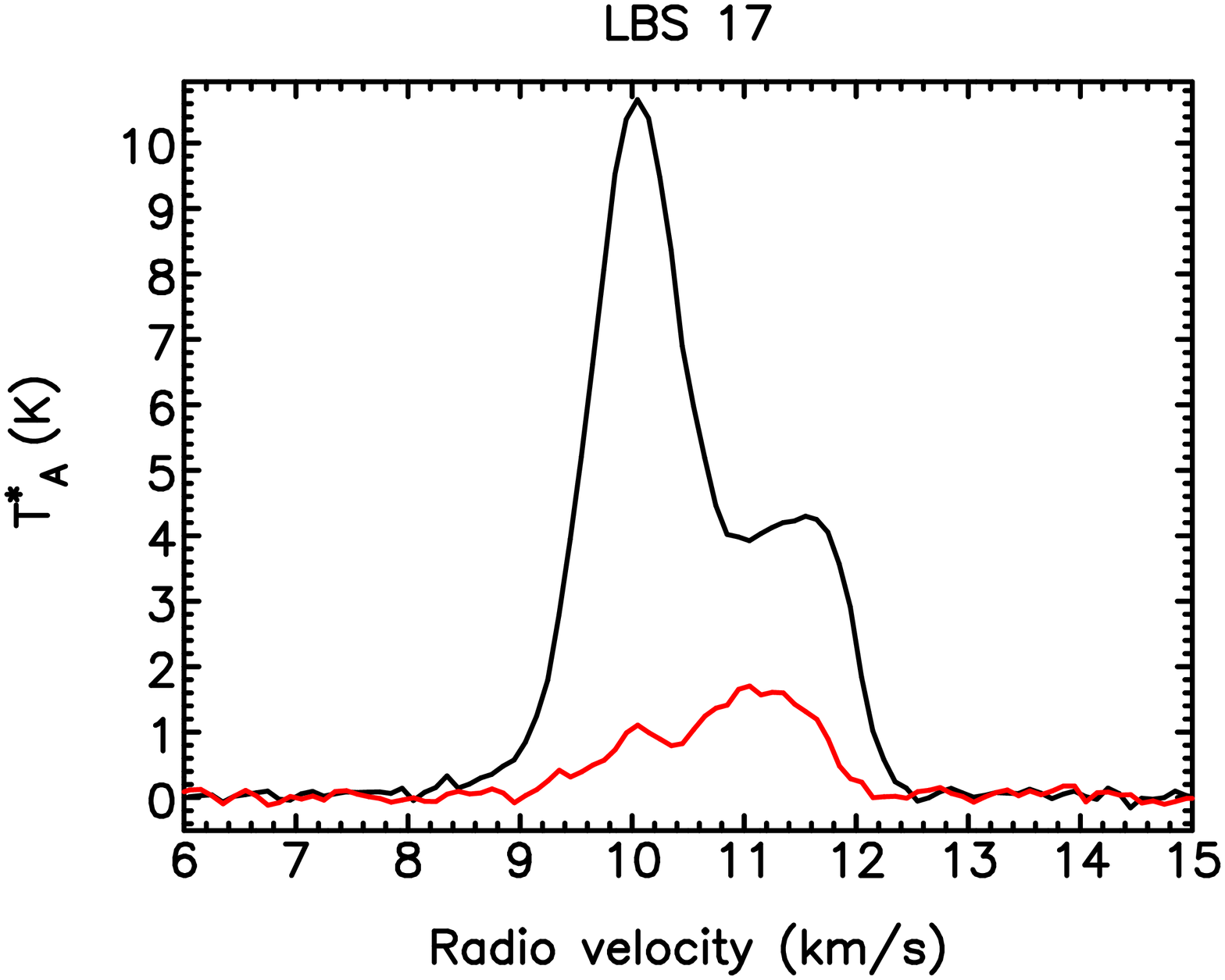}}
\caption{Sample (single-pixel) spectra from different parts of the region. \thirteenCO\ (black line with higher emission) and \CeighteenO\ (red line) spectra from position $05^{\rm h}46^{\rm m}24.6^{\rm s}$, +00$^{\circ}$01\arcmin06\arcsecs in LBS 17 (right) and$05^{\rm h}46^{\rm m}46.2^{\rm s}$, +00$^{\circ}$01\arcmin06\arcsecs in LBS 10 (left) plotted on the same axes.}
\label{spectra}
\end{figure}

\begin{figure}
\centerline{\includegraphics[width=3.4cm,angle=-90]{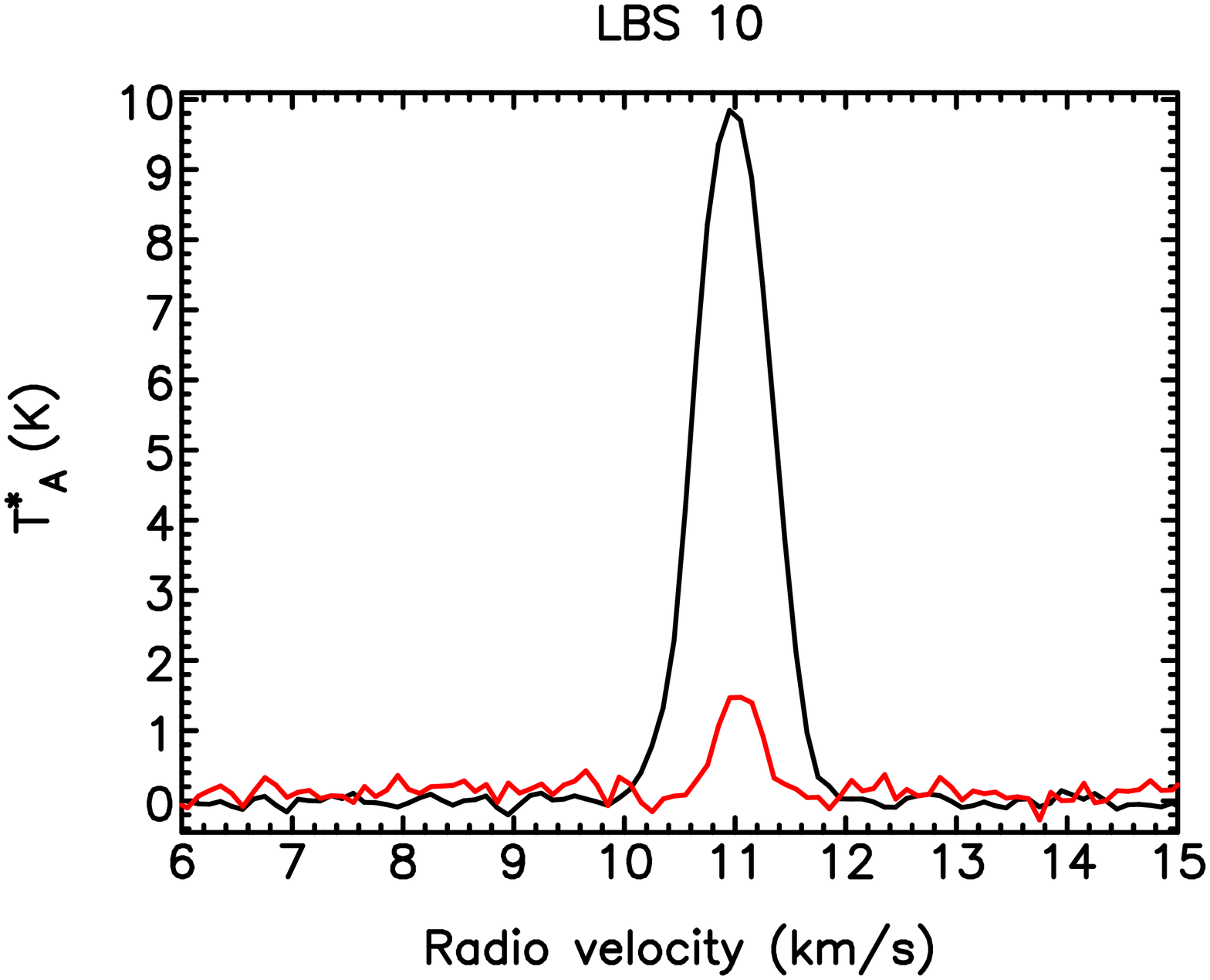}\quad
            \includegraphics[width=3.4cm,angle=-90]{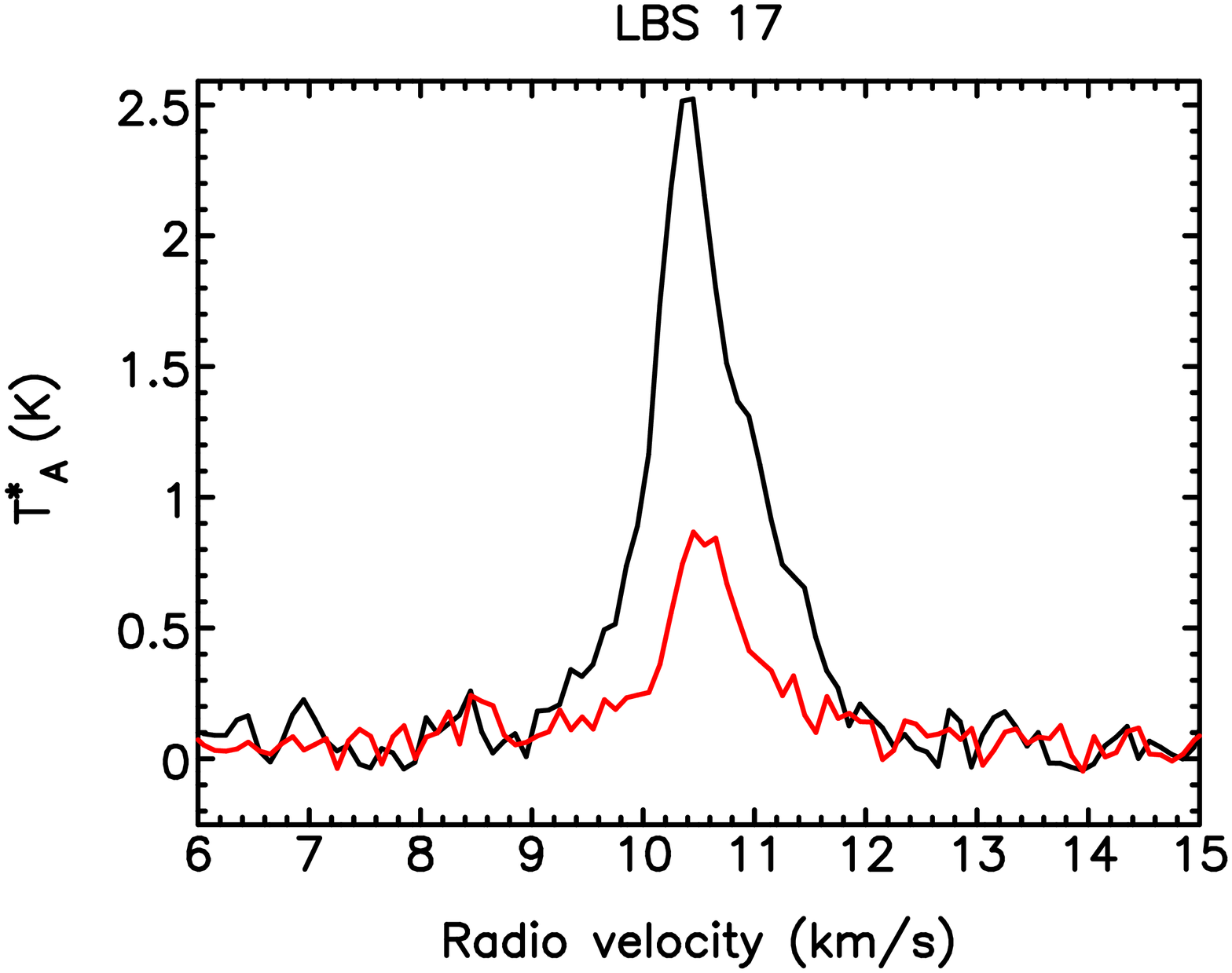}}
\caption{Sample (single-pixel) spectra from different parts of the region. \CeighteenO\ (black line with higher emission) and \HthirteenCOplus\ (red line) spectra from position $05^{\rm h}46^{\rm m}31.3^{\rm s}$, -00$^{\circ}$02\arcmin31.5\arcsecs in LBS 17 (right) and position $05^{\rm h}46^{\rm m}47.8^{\rm s}$, +00$^{\circ}$00\arcmin37.5\arcsecs in LBS 10 (left) plotted on the same axes. The \HthirteenCOplus\ emission is multiplied by 2 times for right-hand plot, and by 5 times in the left hand plot.}
\label{spectratwo}
\end{figure}

Sample \thirteenCO\ and \CeighteenO\ spectra from NGC~2068 can be seen in Figure \ref{spectra}, and differ greatly depending on the part of the region observed --- the spectra in the left and right panels of Figure \ref{spectra} come from the east and west of the region respectively. The \thirteenCO\ emission from the west (corresponding to the LBS 17 clump) has a more complex structure --- a main peak between 9.5 to 10.0 \kms\ and either a lower peak between 10.5 to 11.0 \kms\ or a broad low shoulder that extends out to 12 \kms. Double-peaked structure can arise in an optically thick line as a result of self-absorption or infall, however in this case, the structure is most likely due to multiple velocity components along the line of sight. The optically thinner \CeighteenO\ also shows the same multiple-peaked behaviour (with more distinctly separated peaks that still occur at similar values to the \thirteenCO), demonstrating the presence of multiple components. The emission in the centre and east of the region (corresponding to the LBS 10 clump; seen in the left-hand plot of Figure \ref{spectra}) is much simpler, with both \thirteenCO\ and \CeighteenO\ showing single-peaked spectra, generally peaking between 10 to 11 \kms. The \thirteenCO\ spectra are generally fairly wide, with FWHM values of around 2 \kms; the \CeighteenO\ spectra are narrower with FWHM values of around 1 \kms. The \HthirteenCOplus\ spectra (which come almost exclusively from the LBS 10 clump) are all single-peaked and narrower, generally peaking between 10 to 11 \kms, and have FWHM values of about 0.5 \kms, as can be seen in Figure \ref{spectratwo}.


Figure \ref{ir_overlay} shows that the emission in the west (LBS 17) coincides with an optically-obscured dust lane, while the emission in the centre and east (LBS 10) lies on top of the bright reflection nebula. This leads to the theory that the emission in the west originates from gas and dust in front of the nebula, with the rest of the emission originates from deeper in the cloud, behind the nebula. This theory is supported by polarimetry analysis carried out by \citet{2002ApJ...571..356M}, who in addition, argue that while the emission originates from two distinct regions, these two regions are most likely not completely spatially separated, implying that the filament does not lie in the plane of the sky. \changed{Instead, the dust filament is thought to be twisted with the western edge lying in the foreground of the optical reflection nebula while the eastern edge lies behind it.} This orientation of the filaments helps explain the multiple peaks in the molecular emission --- the lower velocity component (that also generally has higher \antemp\ values and is prevalent in the western part of the map) lies in the foreground and the higher velocity component lies in the background. 

\subsection{High-velocity material}

\begin{figure*}
\centerline{\includegraphics[width=4cm,height=4cm,angle=-90]{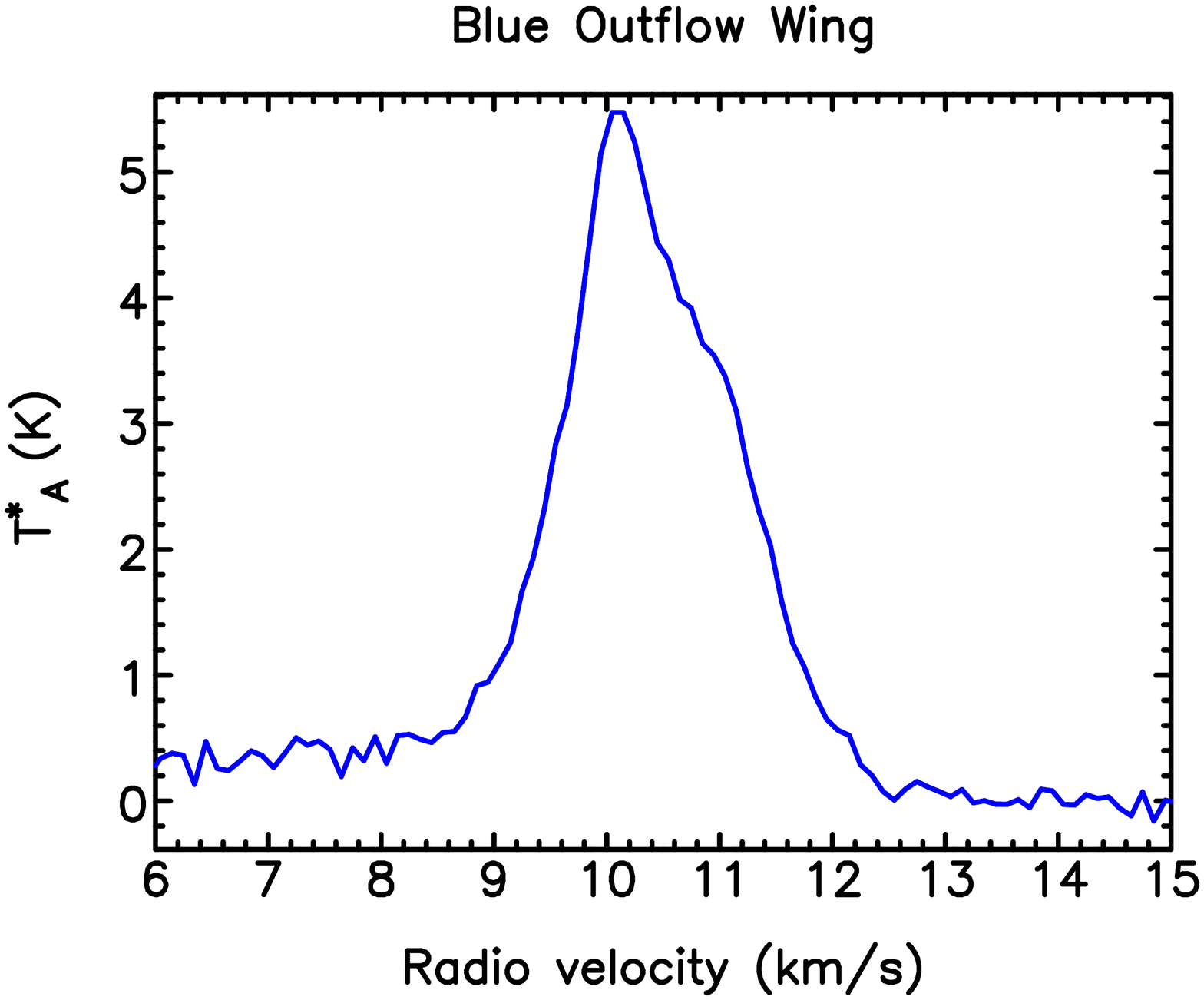}\quad
            \includegraphics[width=4cm,height=4cm,angle=-90]{outflow_cen.ps}\quad
	    \includegraphics[width=4cm,height=4cm,angle=-90]{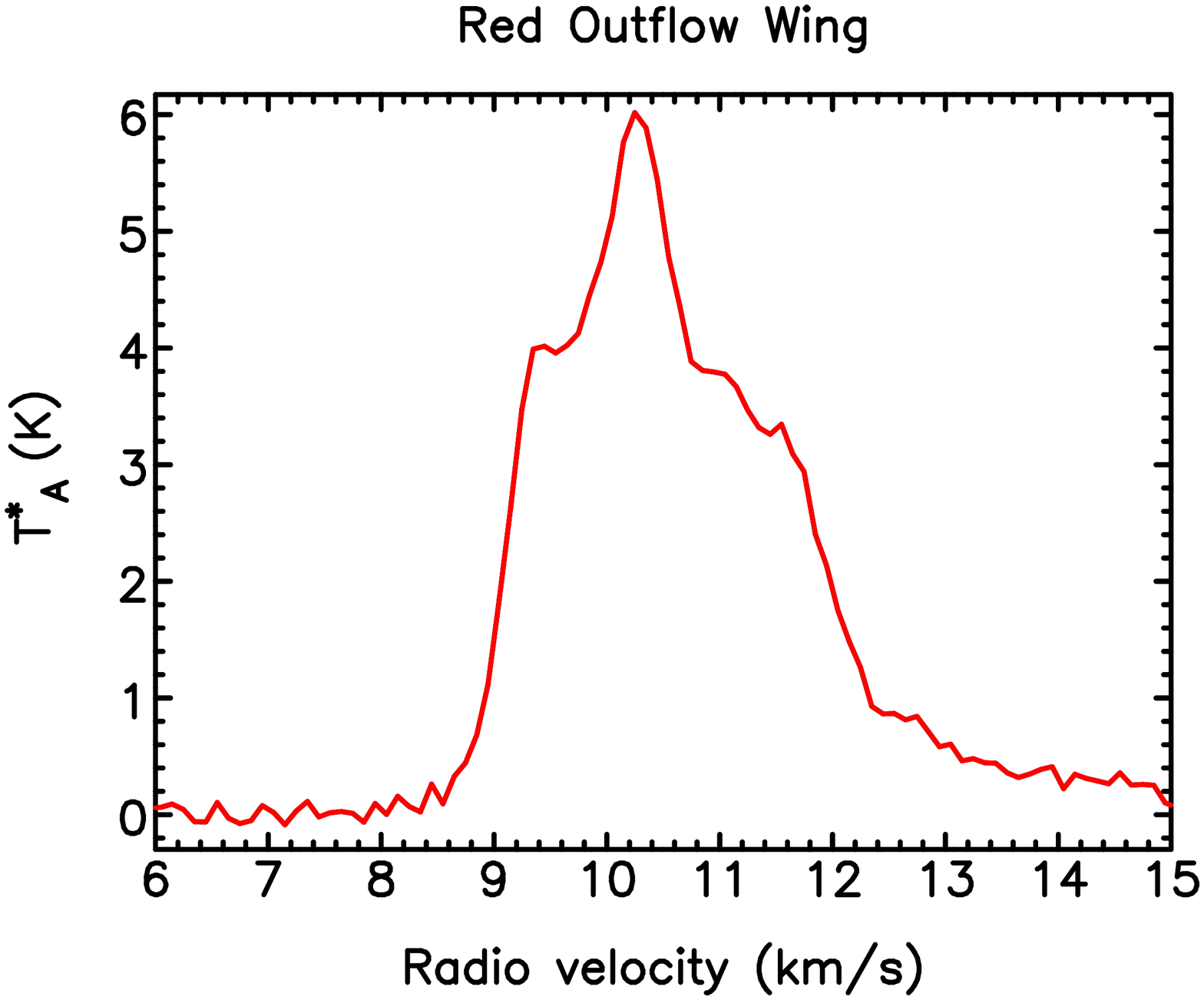}}
\caption{\thirteenCO\ spectra from \changed{3 single pixels around the position of} LBS17-MM9 (a known Class 0 YSO), demonstrating the presence of a strong outflow source. The red and blue spectra (to the left and right) show the outflow wings clearly. \changed{The central pixel still shows residual wings as the outflow is very strong.}}
\label{lbs17mm9outflow}
\end{figure*}

There is evidence of line-wing emission in the \thirteenCO\ data (and to a much lesser extent, also in the \CeighteenO\ data), an example of which can be observed in Figure \ref{lbs17mm9outflow}. In order to capture as much of the outflow as possible without contamination from the central emission peak, we have integrated from 0 to 8.5 \kms\ for blue-shifted emission, and from 12.5 to 20.0 \kms\ for red-shifted emission. This integrated line-wing data is overlaid on SCUBA data in the right-hand panel of Figure \ref{kinematics}, to demonstrate correlation with the M01 SCUBA cores. There are at least 4 separate cases of spatially separated red and blue peaks, which we take to be bipolar outflows.

The isolated bipolar outflow to the west clearly originates from the protostellar core at its centre. This core is identified as LBS17-MM9/OriBsmm51, and has also been studied by \citet{2000MNRAS.313..663G} who concluded that the outflow was driven by a deeply embedded Class 0 YSO. The other outflows in the centre of the map are much harder to disentangle and match to a particular core due to the high density of identified SCUBA cores in the central region. LBS10-MM7/OriBsmm35 and LBS10-MM3/OriBsmm38 also appear to be at the centre of bipolar outflows, have both been identified by NWT as Class I YSOs, and are therefore good outflow source candidates. It is not possible to identify them positively as the driving sources of the outflows, due to the proximity to LBS10-MM5/OriBsmm34 (also identified as a Class I YSO) and LBS10-MM6/OriBsmm36. \citet{2001ApJ...556..215M} mapped the region using \twelveCO\ $J =$ 3 -- 2 data, but were also unable to disentangle the outflows and came to similar conclusions. It is likely that since the cores are so close together, without very high resolution observations (using an interferometer like ALMA for example), it is not possible to pinpoint the exact core driving each outflow. There are patches of both red- and blue-shifted emission in the east that are less well-collimated than the others in the region, that could originate from LBS10-MM9/OriBsmm39 (identified as a Class I YSO). There is also a patch of red-shifted emission in the west that could orginate from LBS17-MM6/OriBsmm47, however there is no corresponding blue-shifted emission, so we cannot postively identify that core as an outflow driving source. 

\section{Clump Decomposition}\label{sec:clfind}
As we are interested in probing the core kinematics and structure, we decompose our \CeighteenO\ and \HthirteenCOplus\ emission into smaller clumps for analysis, as well as for comparison with numerical simulations carried out by others. Both these molecules trace optically thin gas at relatively high densities ($n_{\rm crit}$ \textgreater $10^4$ cm$^{-3}$), and are therefore likely to reveal information about the dense cores and their envelopes/disks. The shapes of the gas clumps produced are also less likely to be affected by high-velocity outflows or background diffuse gas, as would be the case if we were to perform clump decomposition on the \thirteenCO\ emission. 
 
\subsection{Data processing and initial results}
We have used the Starlink {\sc cupid} implementation of the \gclumps\ algorithm, developed by \citet{1990ApJ...356..513S}, to identify Gaussian clumps in our molecular emission. The algorithm fits a triaxial Gaussian profile to the highest peak in the map, then subtracts the fitted profile from the map, iterating until a user-defined termination criteria is reached. We used this method instead of the more common \clfind\ algorithm \citep{1994ApJ...428..693W} as \gclumps\ allows more than one clump to be fitted to a particular peak in the data, and therefore permits clumps to overlap. This is especially useful in the high density centre of the region, as it is not always clear where emission from one core ends and another begins, and the algorithm used must be good at handling blended clumps. 

The termination criteria for the \gclumps\ algorithm were set to be the following: if more than 10 clumps were fitted with peak values less than twice the RMS value; or if 10 consecutive failures occurred in the fitting process. To prevent this, the noisy edges of the cubes had to be removed and the image cubes were cropped in {\sc kappa} to just encompass the main filaments of emissions. Any large-scale ($>$ 2.5\arcmin) background fluctuations in emission were removed using {\sc cupid findback} to smooth the cubes spatially. 

\begin{figure}
\centerline{\includegraphics[width=8.3cm]{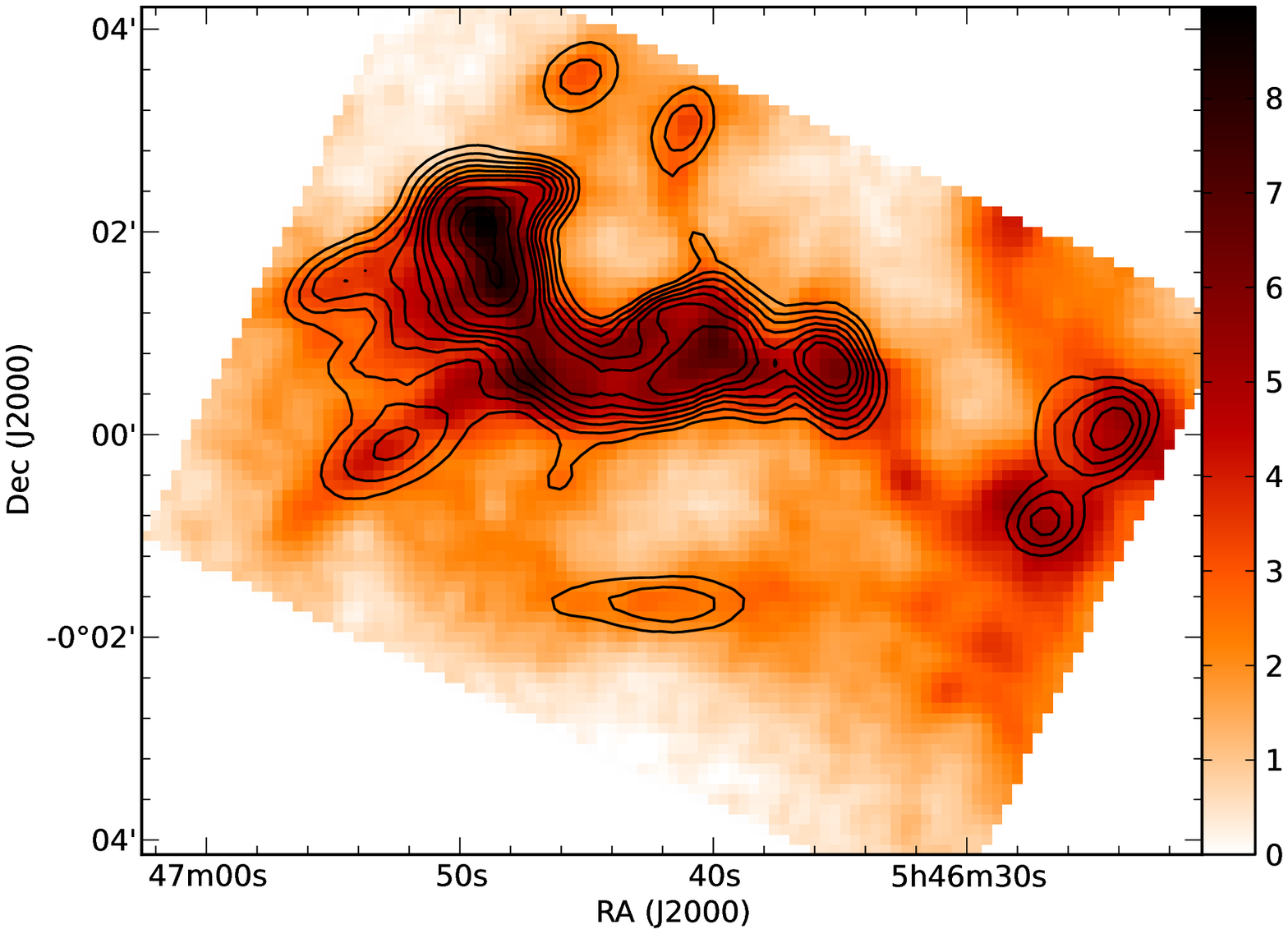}}
\smallskip
\centerline{\includegraphics[width=8.3cm]{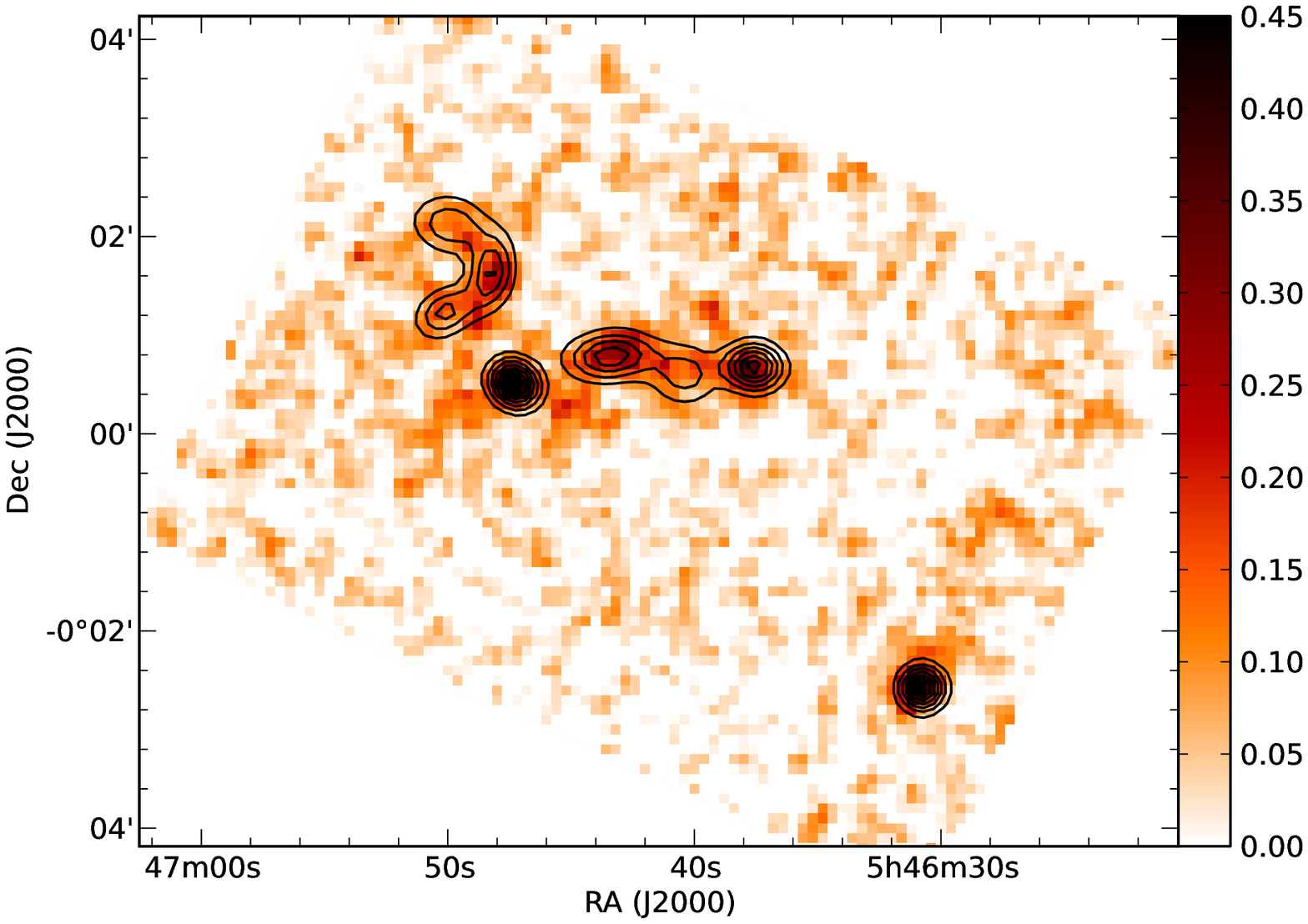}}
\caption{\changed{\gclumps\ fits are overlaid on integrated intensity \CeighteenO\ (top) and \HthirteenCOplus\ (bottom) maps of the observations.} \CeighteenO\ contours are between 1.0 -- 4.0 K~\kms, at 0.5 K~\kms\ intervals, and then between 4.0 -- 10.0 K~\kms, at 1.0 K~\kms\ intervals; \HthirteenCOplus\ contours are between 0.05 -- 0.5 K~\kms, at 0.05 K~\kms\ intervals.}
\label{gclumps_overlay}
\end{figure}

\changed{The \gclumps\ fits ({i.e.} the sum of all the clumps found for each molecule) are plotted over the integrated intensity observations for both molecules in Figure \ref{gclumps_overlay}}. We can see that the \gclumps\ algorithm produces clumps that trace the emission fairly closely (especially in the case of the \HthirteenCOplus), with the fitted cubes containing around 60\% of the original flux. The remaining flux is mainly contained in clumps that are smaller than the effective beamwidth, and are hence discounted in the final analysis. 

\begin{figure}
\centerline{\includegraphics[width=8.0cm]{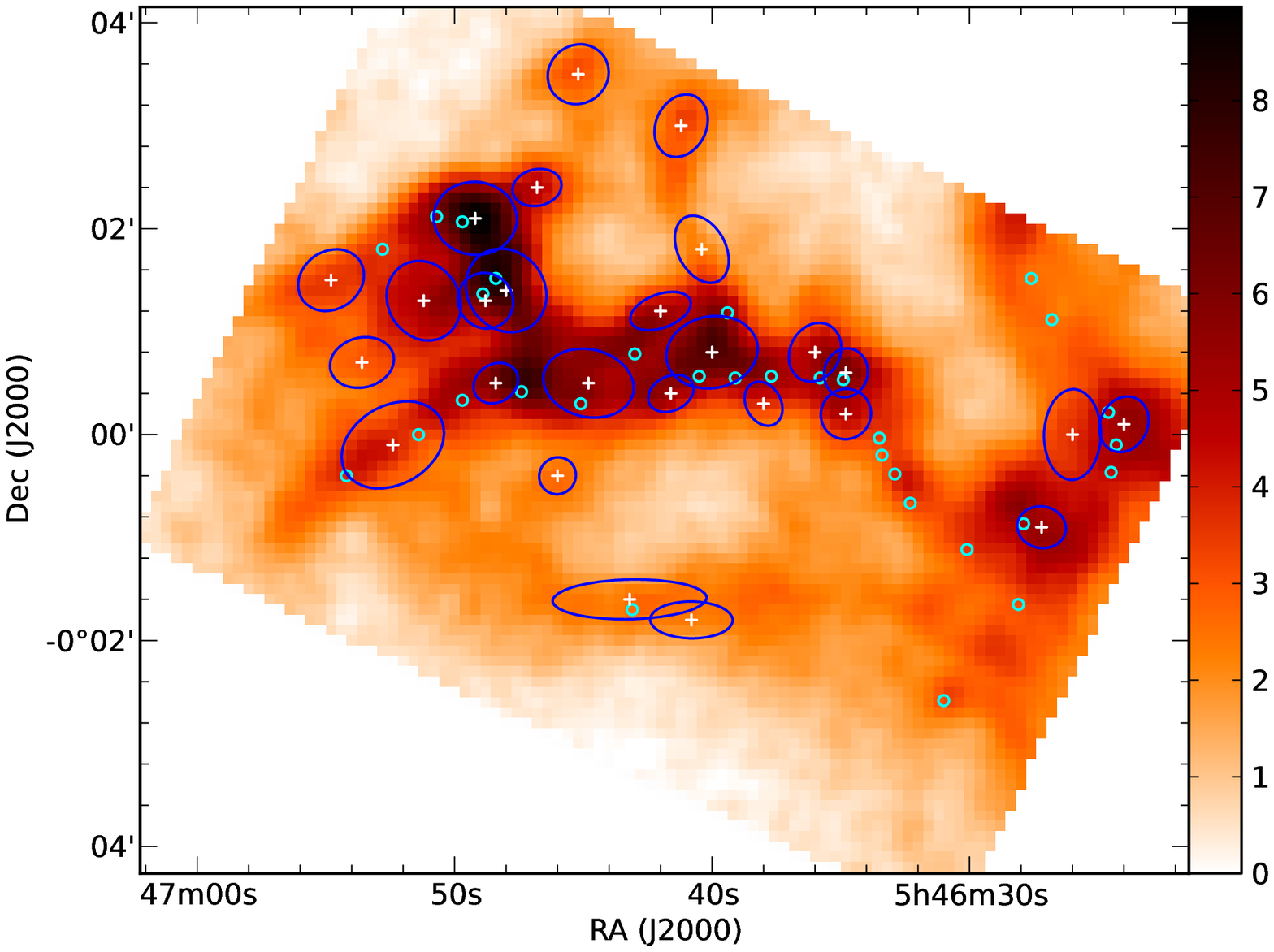}}
\smallskip
\centerline{\includegraphics[width=8.0cm]{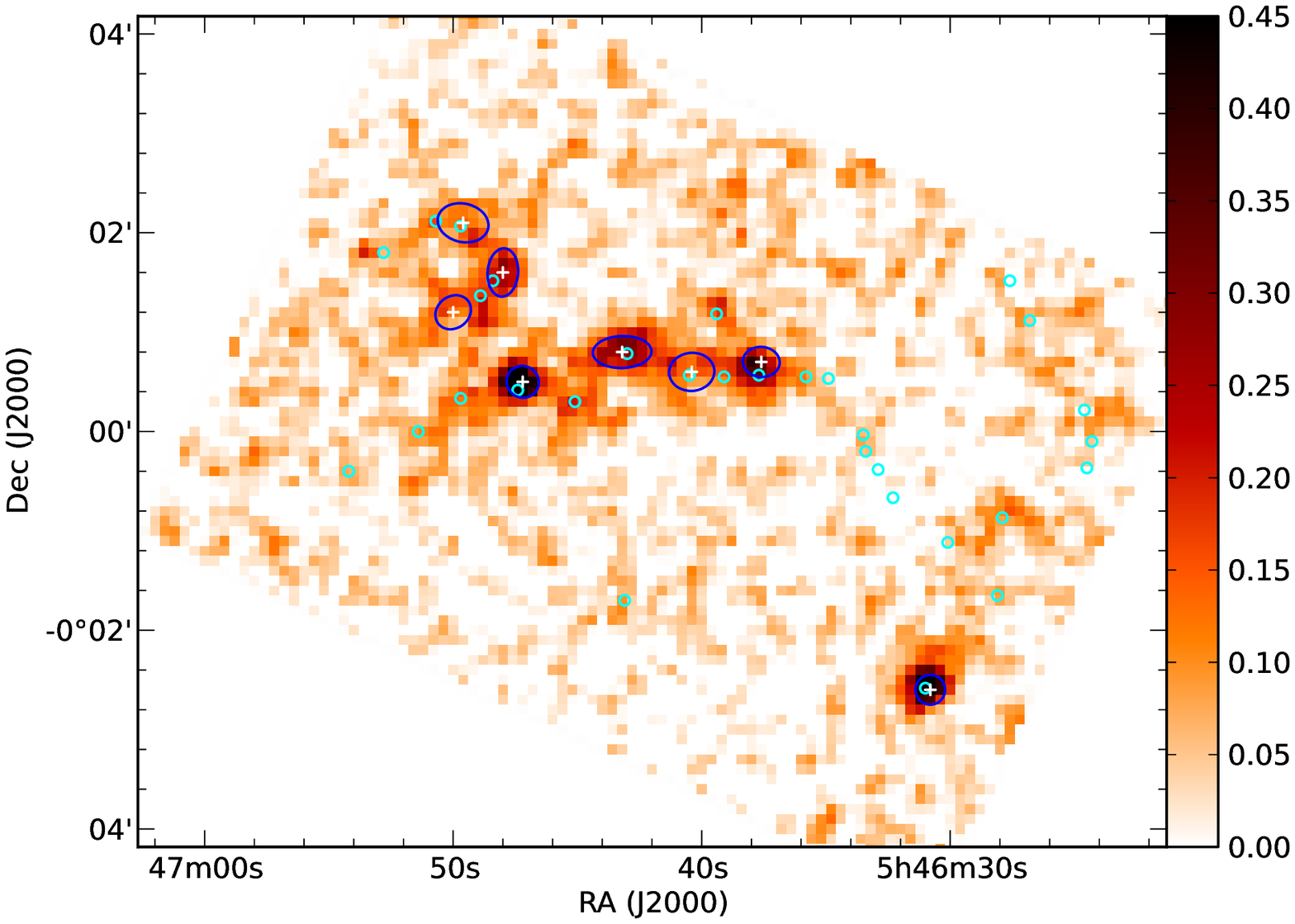}}
\caption{Top: \CeighteenO\ integrated intensity, Bottom: \HthirteenCOplus\ integrated intensity. Ellipses show the FWHM clump extents as determined by \gclumps\ for each molecule, and crosses denote the clump centres. Circles show the positions of the M01 SCUBA cores.}
\label{gcl_ellipses}
\end{figure}

\gclumps\ found 8 \HthirteenCOplus\ clumps and 26 \CeighteenO\ clumps, using the RMS values shown in Table \ref{gcl_numbers}. The ellipses shown in Figure \ref{gcl_ellipses} represent the Gaussian FWHM in the two spatial dimensions, as calculated by the \gclumps\ algorithm. Each clump has been matched by eye with the closest M01 core(s) --- all cores that lie within the ellipse of a clump are matched to that particular clump. All but one of the clump-core matches for \HthirteenCOplus\ produced a one-to-one correspondence, indicating that \HthirteenCOplus\ traces the denser core material; the \CeighteenO\ clumps on the other hand tend to be matched to more than one SCUBA core, as they trace a larger amount of the less dense material surrounding the cores.

\begin{table}
\caption{Results of the \gclumps\ clumpfinding on \CeighteenO\ and \HthirteenCOplus. Column 2: the specified RMS value for each data set, which determines the \gclumps\ termination criteria; Column 3: the total number of clumps found; Column 4: the number of M01 SCUBA cores with a clump match (there were 31 M01 cores in total); and Column 5: the number of clumps without a core match.}
\begin{center}
\begin{tabular}{ccccc}
\hline
Molecule & RMS/~K & Clumps & M01 matches & Noise clumps\\ \hline
\HthirteenCOplus\ & 0.024 & 8 & 8 (26\%) & 1 (13\%)\\
\CeighteenO\ & 0.9 & 26 & 19 (65\%) & 11 (42\%)  \\
\hline
\end{tabular}
\end{center}
\label{gcl_numbers}
\end{table}

\subsection{Comparison with simulations}
\citet{2009MNRAS.396..830S}, hereafter SCB09, developed a gravitational potential clumpfinding algorithm and applied it to simulated data, identifying two different types of p-cores (potential cores): bound p-cores which quickly collapse and form stars that accrete core mass, and composite p-cores which were bound at one point during the simulation but do not necessarily remain bound by the end of the simulation. These p-cores are generally smaller than most dense cores that can be identified observationally, and can only be identified when positions and velocities are known in all three dimensions. This is impossible to replicate observationally, and their closest observational counterparts are probably high-resolution observations of cores in very nearby molecular clouds \citep{2009MNRAS.396..830S}. Nevertheless, we have found it interesting and useful to compare the clumps we identify using more traditional clumpfinding methods, with those identified using the new gravitational potential method. 

\section{Clump properties}\label{sec:clprop}
The \CeighteenO\ is expected to trace gas at lower densities than the \HthirteenCOplus, and the two molecules will therefore probe different regions of star-forming cores. In order to investigate these differences, we examine and compare the properties of the clumps produced for the two molecules. \citet{2009ApJ...691.1560I} also conducted a survey of dense gas cores in Orion B (which included NGC~2068), using \HthirteenCOplus\ $J =$ 1 -- 0 molecular emission, and identified 151 dense cores using the \clfind\ algorithm \citep{1994ApJ...428..693W}. We present a comparison of our clump properties with both SCB09 core properties, and with those cores identified by \citet{2009ApJ...691.1560I} in Table 4.

\subsection{Shapes and Sizes}
The FWHM sizes (in both the spatial dimensions) for each clump were calculated for the two molecules by the \gclumps\ algorithm, and the values were then deconvolved using the effective beamwidth as calculated in Section 2.2. The mean deconvolved aspect ratios of the clumps for both molecules are similar: the value for \HthirteenCOplus\ clumps is $0.80 \pm 0.15$, while that for the \CeighteenO\ clumps is $0.77 \pm 0.16$, where the errors given are the standard deviations\footnote{In all cases, unless otherwise stated, all errors quoted are the standard deviations of the sample}. \changed{The clumps all tend to follow the orientation of the filament they belong to, indicating that they retain the geometry of the filament in which they were formed. As shown in Figure \ref{gcl_ellipses}, for both \CeighteenO\ and \HthirteenCOplus, the major axis is aligned with the filamentary structure seen in the integrated intensity map.}

\begin{figure}
\centerline{\includegraphics[width=4.0cm,height=3.8cm]{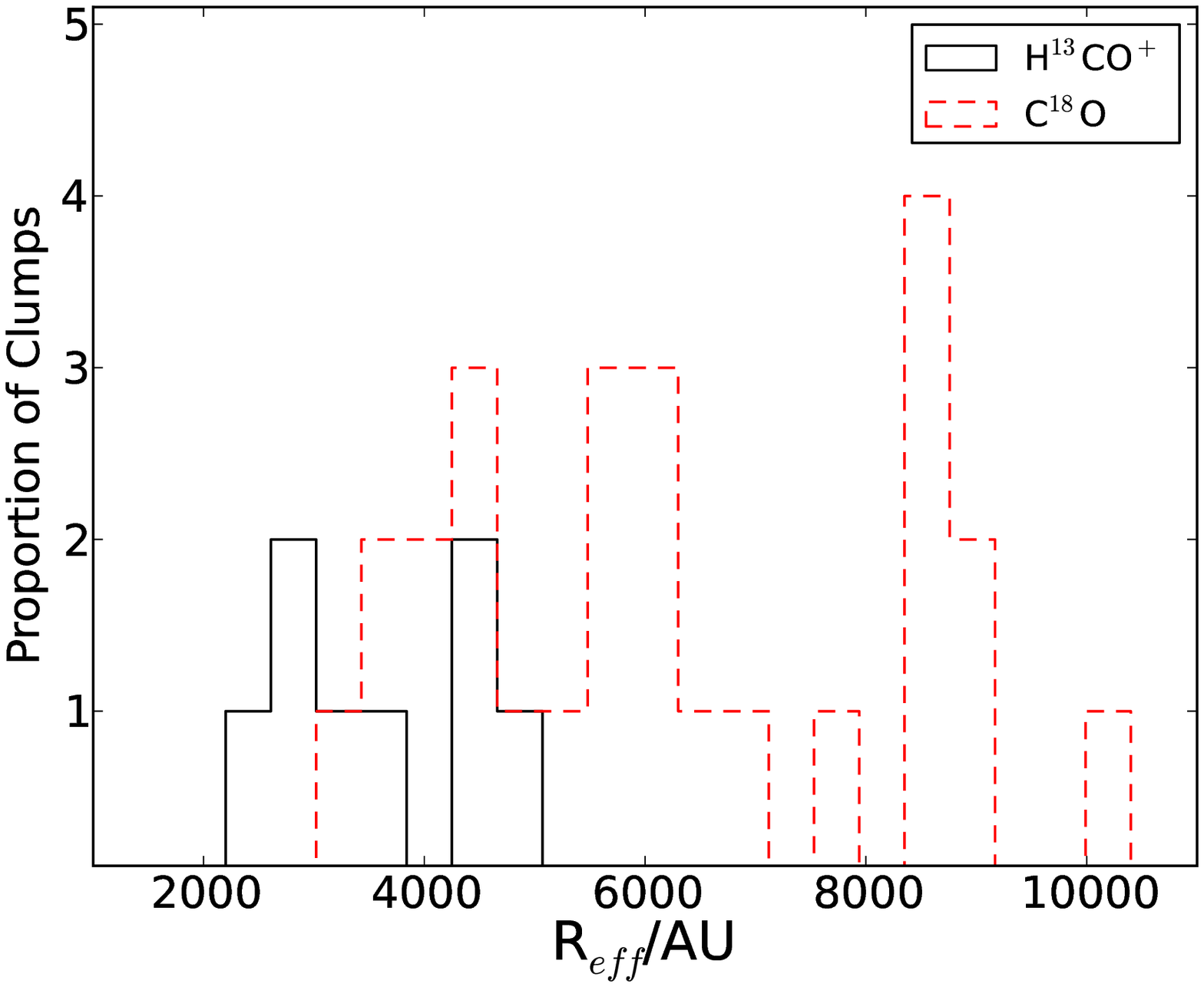}\qquad
	    \includegraphics[width=4.0cm,height=3.8cm]{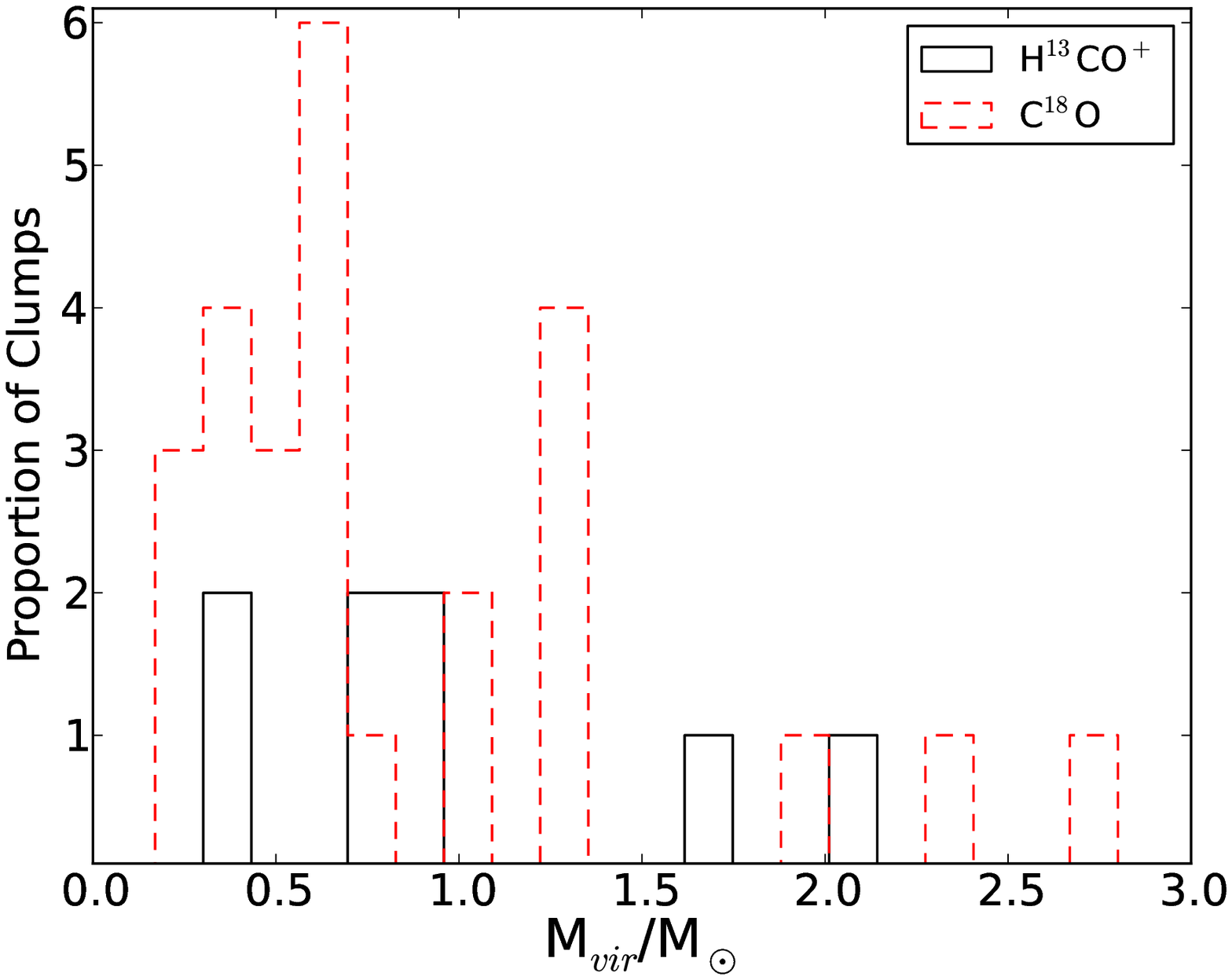}}
\caption{Left: Effective radii for the \HthirteenCOplus\ and \CeighteenO\ clumps;  Right: Calculated virial masses for the \HthirteenCOplus\ and \CeighteenO\ clumps. The \CeighteenO\ clumps have a larger range of radii and masses than the \HthirteenCOplus\ clumps.}
\label{reff_hist}
\end{figure}

The clump effective radii r$_{\rm eff}$ have been calculated by taking the geometric mean of the deconvolved FWHM sizes in the two spatial dimensions. The \HthirteenCOplus\ clumps are on average much smaller than the \CeighteenO\ clumps, as can be seen in the left-hand plot of Figure \ref{reff_hist}. The effective radii of the \HthirteenCOplus\ clumps is $3600 \pm 900$~AU, compared with that of \CeighteenO\ which have effective radii of $6200 \pm 2000$~AU. The M01 SCUBA cores have a deconvolved radii of $\sim 2500$AU \citep{2001A&A...372L..41M}, which implies roughly a one-to-one correspondence with the \HthirteenCOplus\ clumps, while each \CeighteenO\ clump will encompass more than one M01 core. Comparing our clumps with the bound and composite p-cores of SCB09 (which have effective radii of 2400 AU and 3700~AU respectively), we find that the \HthirteenCOplus\ clumps have $r_{\rm eff}$ comparable to the average p-core radius, and are therefore more likely to correspond to individual cores. The \CeighteenO\ clumps on the other hand are much bigger and encompass \changed{multiple p-cores; these are expected to correspond to the bound p-cores which have masses that are about half that of the \CeighteenO\ clumps on average (see Table \ref{allcomp}}.

\subsection{Linewidths}
The average linewidth for each clump calculated by \gclumps --- the FWHM of the Gaussian fit to the average spectrum for the clump --- is generally smaller than the linewidth at the clump central position. In fact, this discrepancy is so much greater for the \CeighteenO\ clumps than for the \HthirteenCOplus\ clumps that the clump average linewidths for \CeighteenO\ are smaller than those of \HthirteenCOplus: $0.18 \pm 0.05$ \kms\ for \CeighteenO\ compared to $0.24 \pm 0.08$ \kms\ for \HthirteenCOplus. However, when a Gaussian fit to the spectra at the clump centre is performed, an average value of 0.28 $\pm$ 0.02 \kms\ for \CeighteenO\ and $0.27 \pm 0.04$ \kms\ for \HthirteenCOplus\ is obtained. These latter values of \CeighteenO\ are a better reflection of the actual observed spectra of the molecules (see Figure \ref{spectratwo}), where the \CeighteenO\ emission line is visibly wider than that of \HthirteenCOplus. \changed{This result appears to contradict the well-known result that the linewidth increases with the size of region traced \citep{1981MNRAS.194..809L}. However, this can be explained by the fitting mechanism in \gclumps: the central peaks are generally well-defined and easy to fit; but at the clump edges, the emission decreases and the fits are made to peaks that have much lower SNR and tend to be much narrower. Therefore the average spectra produced have a significant contribution from very narrow spectra, causing them to be narrower than the peak spectrum. This effect is more noticeable for the \CeighteenO\ clumps than for the \HthirteenCOplus\ clumps, because the latter tend to have a much sharper cutoff as the emission is very concentrated.} Therefore, in all subsequent calculations, to standardise the comparisons and to avoid biasing the spectra due to noise contributions, we will use the clump centre linewidths \sigmaoned\ and central velocities $v_{\rm c}$, rather than the average values produced by \gclumps.

\begin{figure}
\centerline{\includegraphics[width=3.9cm,height=3.8cm]{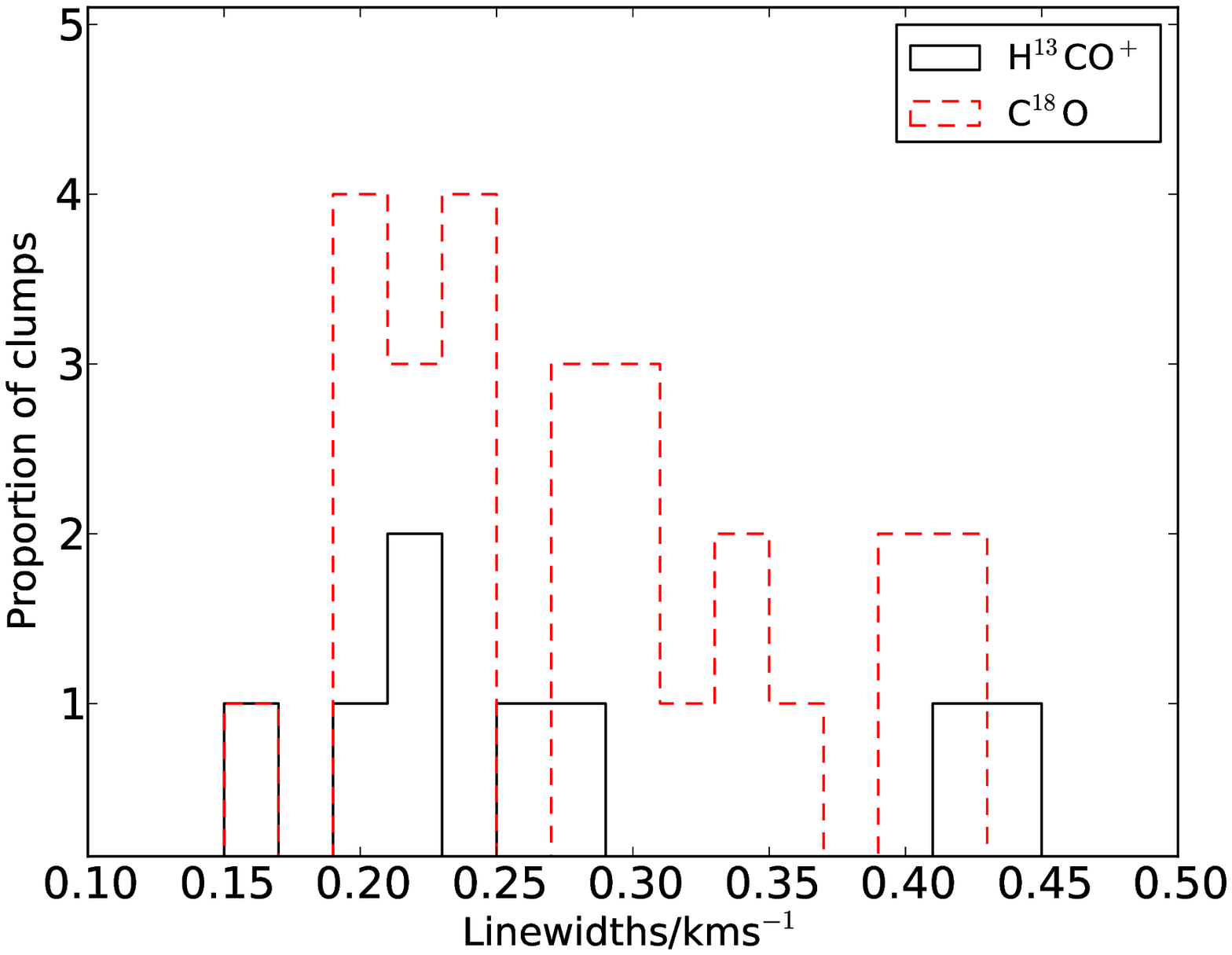}\qquad
	    \includegraphics[width=4.1cm,height=3.8cm]{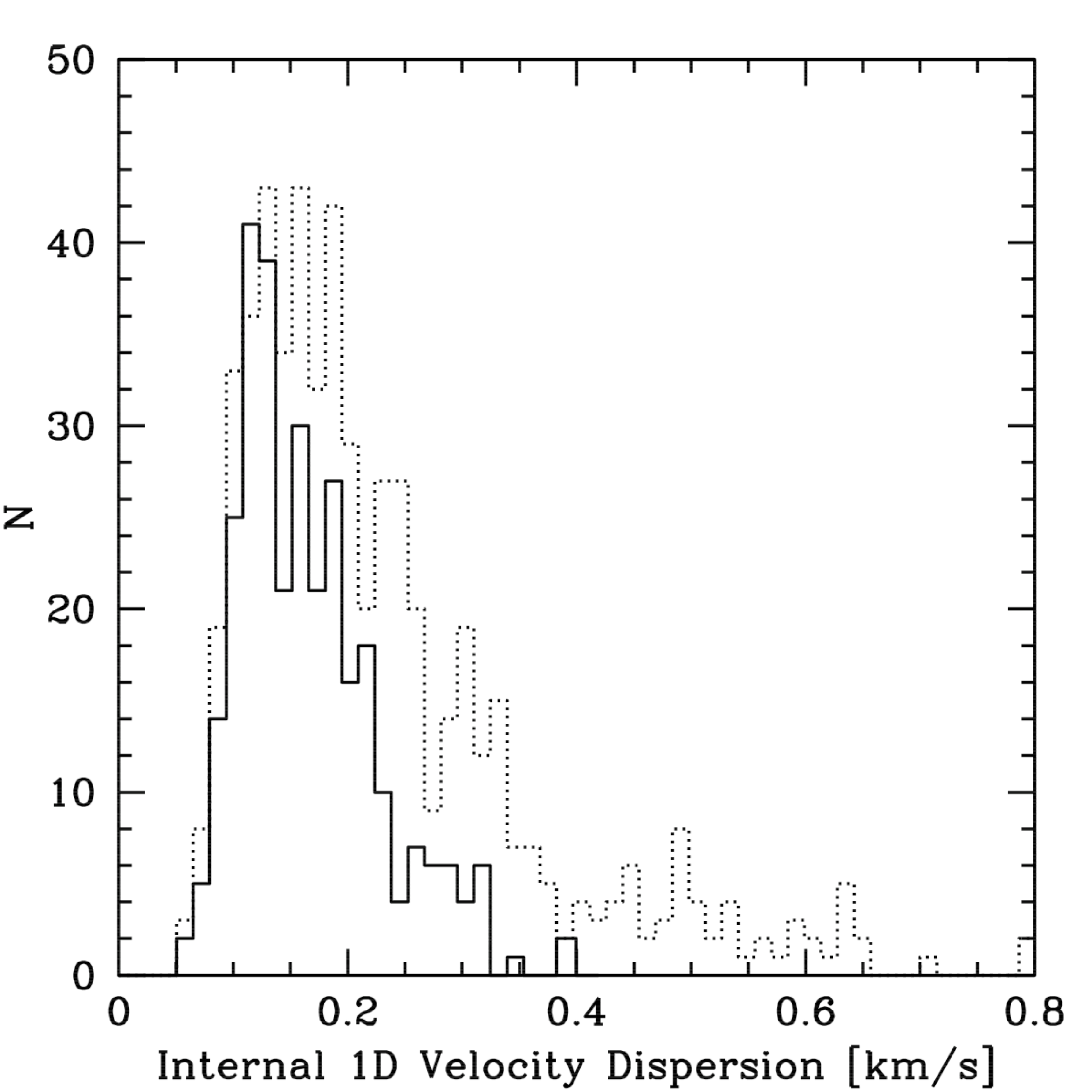}}
\caption{Left: Histogram showing the linewidths of the clumps in the \CeighteenO\ (dashed line) and \HthirteenCOplus\ (solid line) datasets. Right: Histogram showing the 1D internal velocity dispersions of the p-cores in the bound (solid line) and composite (dotted line) datasets, taken from SCB09.}
\label{cl_lw}
\end{figure}

The linewidths are shown in Figure \ref{cl_lw}, with the internal 1D velocity dispersion of the p-cores from SCB09 for comparison. We see that the distribution of both sets of clumps peaks at a much higher value than the SCB09 p-cores --- SCB09 calculate \sigmaoned\ values of 0.16 \kms\ and 0.23 \kms\ for bound and composite p-cores respectively. This is to be expected for our \CeighteenO\ clumps as they are much larger than the SCB09 cores, and trace more of the turbulent diffuse gas surrounding star-forming cores. The \HthirteenCOplus\ clumps trace very dense gas, and would be expected to have much smaller linewidths. However, over half of the clumps are associated with centrally condensed protostellar objects; these clumps exhibit higher linewidths than those that are starless and therefore increase the average clump linewidth. The average linewidth of those \HthirteenCOplus\ clumps not associated with a protostar is 0.17 $\pm$ 0.05 \kms, which matches the values obtained for the SCB09 cores much better. \changed{It should be noted that the thermal width of \HthirteenCOplus\ at 16~K is 0.07 \kms, making the thermal contribution to the linewidth less than 10\%.} 

We analyse the clump linewidths in greater detail, and also discuss our results in relation to theories of star formation, as well as timescales and size scales of turbulence, in Section 5.6.

\subsection{Virial Masses}
The virial mass of a spherically symmetric clump can be calculated using the following equation \citep{1988ApJ...333..821M}:
\begin{equation}
M_{\rm vir} = \frac{k_1 r_{\rm eff}\sigma_{3\rm D}^2}{G},
\label{mvir}
\end{equation}
where $\sigma_{3\rm D}^2 = 3\sigma_{1\rm D}^2$, r$_{eff}$ is the effective radius of the clump (as discussed in Section 5.1), and G is the gravitational constant; $k_1$ is a constant whose exact value depends on the density distribution of the clump as a function of distance from the clump centre. For simplicity we assume that the clumps follow $\rho \propto r^{-2}$ density distributions, yielding $k_1 = 1$.

The virial mass of the \HthirteenCOplus\ clumps varies beween 0.3 -- 2.0 \Msun, while that of the \CeighteenO\ clumps has a range of 0.2 -- 2.7 \Msun\ (see Figure \ref{reff_hist}). In general, the average \CeighteenO\ virial mass of 0.86 $\pm$ 0.13 \Msun\ is similar to that of the \HthirteenCOplus\ clumps (0.97 $\pm$ 0.21 \Msun).  

\subsection{Gas Masses}

\begin{table}
\caption{Various properties of \HthirteenCOplus\ and \CeighteenO\ used in this paper to calculate column densities and gas masses. \changed{Values of $T_0$, the energy of the lowest state, in K,} are obtained from the Leiden Atomic and Molecular Database (LAMBA) \citep{2005A&A...432..369S}. The relative abundances X$_{\rm gas}$ were obtained from \citet{1982ApJ...262..590F} and \citet{2009ApJ...691.1560I}.}
\begin{center}
\begin{tabular}{ccc}
\hline
Molecule & $T_0$ /K & X$_{\rm gas}$\\ \hline
\HthirteenCOplus\ & 4.16 & $4.3 \times 10^{-11}$\\
\CeighteenO\ & 5.27 & $1.7 \times 10^{-7}$\\
\hline
\end{tabular}
\end{center}
\label{coldenprops}
\end{table}

The gas mass of the clumps can be calculated from the column density of the gas, assuming Local Thermal Equilibrium (LTE). Following \citet{2000..RW..book} and assuming values for the different molecules as presented in Table \ref{coldenprops}:

\begin{equation}
N_{\rm H^{13}CO^+}=5.70\times 10^{13}{\rm m}^{-2} \frac{T_{\rm ex}}{\exp({-41.7 K/T_{\rm ex}})} \int{T_A^*{\rm d}v}.
\end{equation}
%
\begin{equation}
N_{\rm C^{18}O}=8.70\times 10^{16}{\rm m}^{-2} \frac{T_{\rm ex}}{\exp({-31.7 K/T_{\rm ex}})} \int{T_A^*{\rm d}v}.
\label{cocol}
\end{equation}
\changed{The output from {\sc cupid} --- the `clump sum' --- is the sum of the data enclosed within the \gclumps\ ellipse (defined by the Gaussian FWHM).} The integrated intensity $\int{T_A^*{\rm d}v}$ is obtained by multiplying this clump sum by the velocity resolution of the cubes (0.1 \kms). In both cases, we take the excitation temperature to be $16.1 \pm 4.7$ K, the average value calculated for the entire region by \citet{2009ApJ...691.1560I}, using NH$_3$ transitions. Using a single temperature to represent all the different cores will introduce a certain inaccuracy to the gas masses calculated, as it will not necessarily reflect the different conditions within each clump. However, it is the least biased method we could choose and allows consistent comparisons for all clumps identified. Additionally, our most optically thick molecule \thirteenCO\ is only barely optically thick (with opacity values of $\tau \sim$ 0.9), and will not give accurate excitation temperatures, tending towards underestimation; we have therefore chosen not to use excitation temperatures calculated from \thirteenCO.

We obtain the masses of the clumps by substituting representative values for the two molecules into the following equation:
\begin{equation}
M_{\rm gas}=\rm D^2(\Delta \alpha \Delta \beta)\mu_{\rm H2} m_{\rm H2} X^{-1}_{\rm gas} N_{\rm gas}
\label{mass}
\end{equation}
We take the distance to the cloud $D = 415$ pc and adopt a mean molecular weight per H$_2$ molecule of $\mu_{\rm H2}=2.72$ to include helium; we also use a pixel size ($\Delta \alpha$ and $\Delta \beta$) of 6\arcsecs\ and take the abundances of \HthirteenCOplus\ and \CeighteenO\ relative to hydrogen ($X_{\rm H^{13}CO^+}$ and $X_{\rm C^{18}O}$) to be as shown in Table \ref{coldenprops}.

The values of the gas masses calculated using this method range from 0.5 to 1.4 \Msun\ for \HthirteenCOplus, and from 0.6 to 12 \Msun\ for \CeighteenO. As expected, the \CeighteenO\ clumps contain more gas mass due to their much larger sizes, having an average mass of 3.7 $\pm$ 3.0 \Msun\ compared to the average mass of 0.92 $\pm$ 0.24 \Msun\ calculated for \HthirteenCOplus. 

\begin{table*}
\caption{Comparison of results from this paper with referenced literature values. The typical clump mass values used in the case of this work, are the gas masses of the clumps, calculated from the column density values. The errors on the last digit are given in brackets for the \sigmaoned\ values.}
\centering
\begin{tabular}{cccccc}
\hline
Author & Region & Clump type & Typical Clump Size/~AU & Typical Clump Mass/\Msun\ & Mean \sigmaoned/\kms\ \\ \hline
Ikeda et. al (2009) & Orion A & \HthirteenCOplus\ (J=1-0) & 2.9 $\times$ 10$^4$ & 2.1 -- 81 & 0.52\\
Smith et. al.(2009) & simulated & p-cores (bound) & 2.4 $\times$ 10$^3$ & 0.2 -- 5 & 0.16\\
Smith et. al.(2009) & simulated & p-cores (composite) & 3.7 $\times$ 10$^3$ & 0.2 -- 10 & 0.23\\[3pt]
This work & Orion B & \HthirteenCOplus\ (J=4-3) & 3.6 $\times$ 10$^3$ & 0.5 -- 1.4 & 0.27(4)\\
This work & Orion B &  \CeighteenO\ (J=3-2) & 6.2 $\times$ 10$^3$ & 0.6 -- 12 & 0.28(2)\\
\hline
\label{allcomp}
\end{tabular}
\end{table*}

\begin{figure}
\centerline{\includegraphics[width=8.5cm]{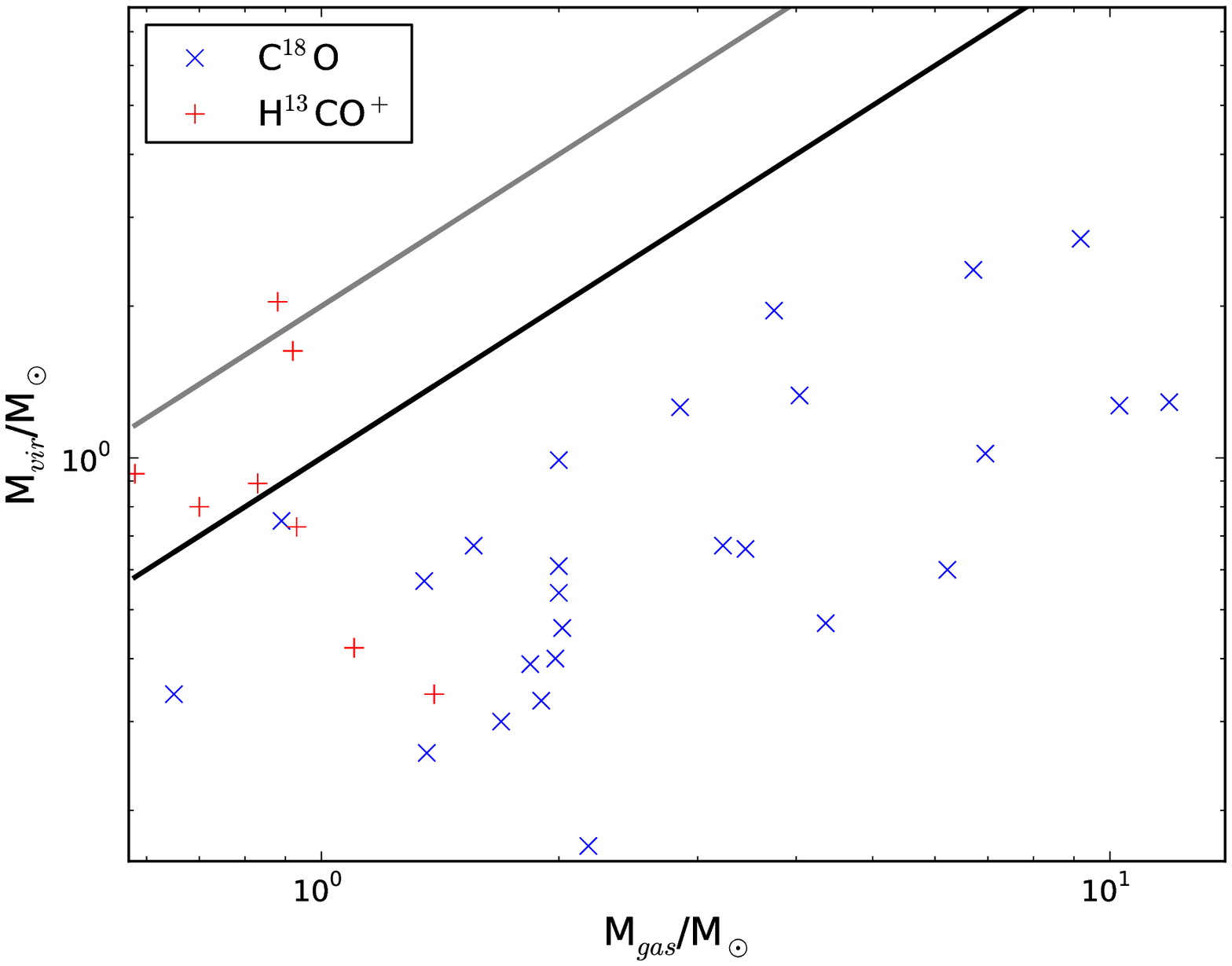}}
\smallskip
\centerline{\includegraphics[width=8.5cm]{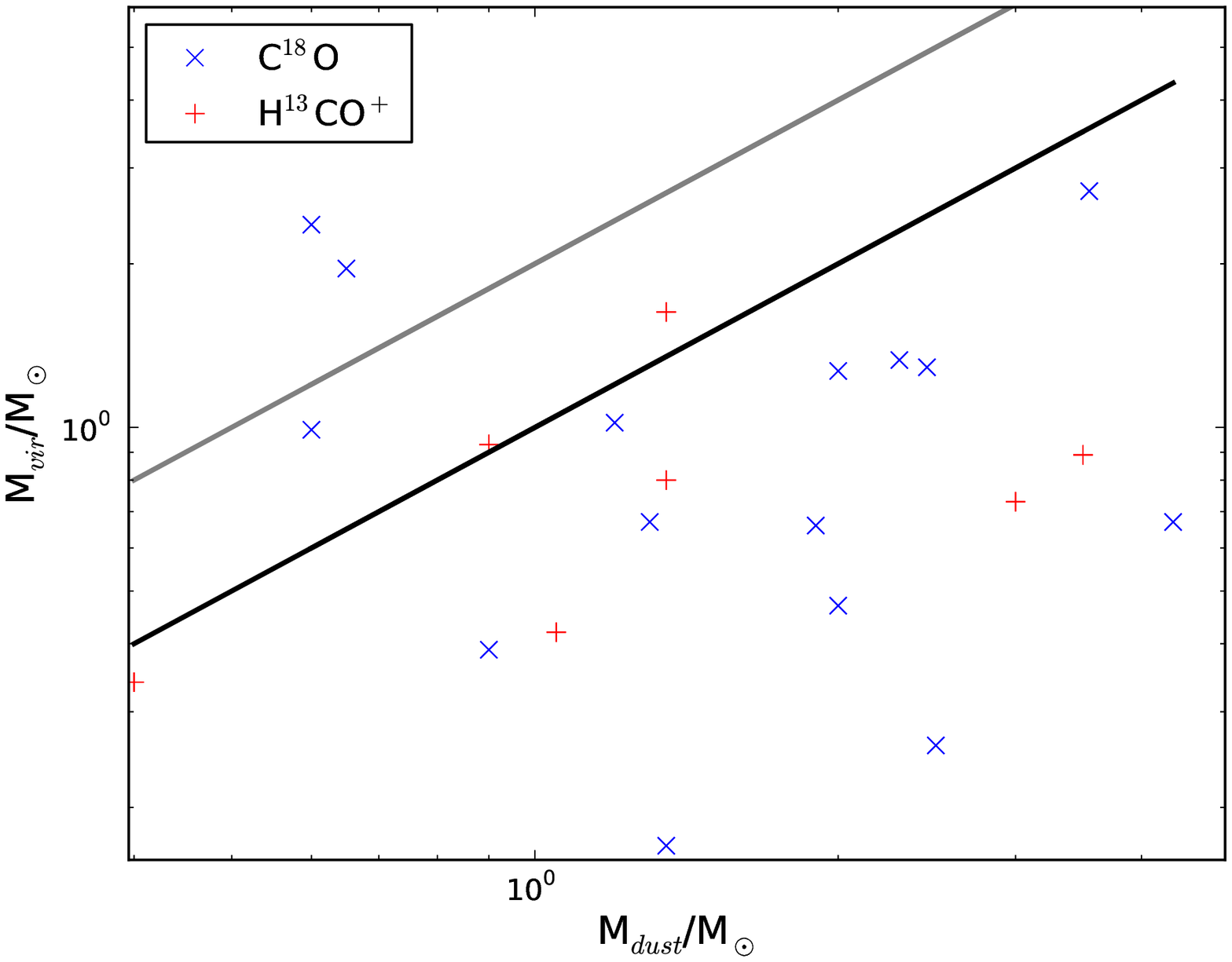}}
\caption{Plots of (Top) $M_{\rm vir}$ against $M_{\rm gas}$ and (Bottom) $M_{\rm vir}$ against $M_{\rm dust}$, for both \HthirteenCOplus\ (red) and \CeighteenO\ (blue). The $M_{\rm vir} = 2M_{\rm gas}$ (bound) and $M_{\rm vir} = M_{\rm gas}$ (equipartition) lines are also indicated using black and grey lines respectively.}
\label{ratioplots}
\end{figure}

\subsection{Bound Ratio}
One method of separating and identifying clumps that are prestellar and likely to collapse to form protostars rather than transient structures, is to determine whether they are bound. A fairly straightforward method of doing this is by comparing their virial masses (a measure of their internal energy) with their gas or dust masses (a measure of the potential energy in the clump). Clumps are generally considered to be bound if $M_{\rm vir} \le 2M_{\rm gas}$; however, the lower limit for equipartition, where the clumps are confined but not necessarily self-gravitating, is taken to be $M_{\rm vir} = M_{\rm gas}$ \citep{1992ApJ...395..140B}. 

Figure \ref{ratioplots} (top) shows a plot of clump virial mass against gas mass, where all of the \CeighteenO\ clumps and a third of the \HthirteenCOplus\ clumps lie to the right of the `bound' line, and all but one of the \HthirteenCOplus\ clumps are at least in equipartition. This implies that the larger \CeighteenO\ clumps are more bound than the smaller \HthirteenCOplus\ clumps. \HthirteenCOplus\ traces very dense gas and we would expect the structures identified to be more tightly bound than the lower density structures traced by \CeighteenO, making our result appear contradictory. For comparison, we also plot clump virial masses against their dust masses (see Figure \ref{ratioplots}, bottom), which are calculated from the dust masses of the corresponding M01 SCUBA cores. If (as is usually the case for \CeighteenO) more than one M01 core corresponds to a particular clump, then the core masses are summed. (Those gas clumps without an M01 core match are omitted in this part of the analysis.) It is reassuring to see that the majority of the \HthirteenCOplus\ clumps now lie to the right of the `bound' line, and that the only clumps lying to the left of the equipartition line (and therefore unbound) are \CeighteenO\ clumps. 

The dust masses seem to give a more accurate reflection of the mass within the \HthirteenCOplus\ clumps probably because they are of similar sizes, while giving an underestimation of the mass in the \CeighteenO\ clumps, which are much bigger and encompass more of the envelope material surrounding the dense cores. \changed{However, there are sources of error (for example, spatial filtering in dual-beam switch observing mode) that could bias the dust masses towards lower values. In addition, there are intrinsic sources of error and uncertainty present, such as dust emissivity and the dust/gas mass ratio, as well as contamination from molecular gas like \twelveCO, that could affect dust mass estimates.}

One of the reasons for the \CeighteenO\ clumps appearing more bound is an overestimation of the gas mass. \CeighteenO\ has a lower critical density than \HthirteenCOplus\ and therefore will exhibit a contribution from the less dense surrounding envelopes of pre- and protostellar cores. This would account for the much larger sizes and higher gas masses of \CeighteenO\ over \HthirteenCOplus\ --- the \HthirteenCOplus\ clumps represent the denser core regions while the \CeighteenO\ clumps represent both core envelopes and centres. In addition, the \HthirteenCOplus\ may have been sub-thermally excited which would result in an underestimation of the column density and hence the mass present in the clump --- this theory is supported by the fact that a large proportion of the \HthirteenCOplus\ clumps have higher dust masses than gas masses.

There are also many sources of error present in the calculation of gas masses, the most significant of which is the value of the abundance ratios of \HthirteenCOplus\ and \CeighteenO, which can cause the calculated gas masses to vary by a factor of 3. 

Finally, the comparison of virial and gas/dust masses merely tells us if the material in the clump is bound with respect to itself, but does not tell us if the clump is bound with respect to the surrounding gas environment. There are other (less easy-to-quantify) methods of confinement for clumps ({e.g.} magnetic fields, external pressure) that could mean that a clump is in fact a bound (and hence non-transient) structure. In fact, converting a linewidth straight to a virial mass without accounting for surface pressure will overestimate the enclosed mass of the core and hence make it appear less bound than it actually is (for example, see \citet{2007ApJ...661..262D}).

\subsubsection{Overall region boundness}
Although the comparison of virial mass with gas mass may not be the most suitable method for determining if small sub-structures within molecular cloud regions are bound, it should still give a good idea of the condition of the region as a whole. The \thirteenCO\ emission lines are resolved across the whole region, and will therefore give a good estimate of the virial mass of the region; the \CeighteenO\ emission also covers the majority of the region, is optically thin throughout (unlike the \thirteenCO), and should give a good estimate of the amount of gas contained within the region. We can thus calculate the virial and gas mass of the NGC~2068 region, and determine whether it is bound as a whole, using a similar method to that described previously. Assuming that the region is spherically symmetric, we therefore calculate the virial mass from: 
\begin{equation}
M_{\rm vir} = \frac{3r_{\rm cond} \sigma^2_{\rm tot}}{G}.
\end{equation}
The value of $r_{\rm cond}$ --- the radius of the NGC~2068 region --- is taken to be 4\arcmin\ (which translates to $\sim$0.5 pc at a distance of 415 pc); and $\sigma_{\rm tot} = 0.61$ \kms, as obtained from the dispersion of \thirteenCO\ line centre velocities over the region. Substituting these values, $M_{\rm vir}$ is approximately 130 \Msun. 

Using Equations \ref{cocol} and \ref{mass}, we calculate $M_{\rm gas}$ of approximately 230 \Msun\ from the \CeighteenO\ emission, which is almost twice the calculated virial mass. This indicates that the cloud is most likely in the process of global collapse, unless additional support ({e.g.} magnetic fields) were present. \changed{In order to provide this support, the magnetic field strength would have to be $\sim$ 59$\mu$G (following virial equilibrium calculations by \citet{1992ApJ...395..140B}), which is similar to the value of $\sim$ 36$\mu$G they have measured in Orion B. We should not however place too much weight on this value due to the complex dynamics and irregular shape of the NGC~2068 cloud.}


\subsection{Velocity dispersions}
The kinematics of fragmenting cores can be used to investigate the initial conditions of star formation. An examination of the internal velocity dispersions of these cores tells us the magnitude of turbulence present within the cores on small scales \citep{2007ApJ...668.1042K}. The relative movement between \emph{cores}, as well as between the cores and their surrounding gas, could shed light on the timescale of turbulent decay present in the region --- whether the turbulence decays on dynamical timescales, or is driven and lasts for longer timescales. \citet{2008AJ....136..404O} found that comparing the core-to-core velocity dispersions of protostellar and starless cores allowed them to differentiate between their driven and decaying turbulence simulations. 

%
\subsubsection{Clump internal velocity dispersions}
The \sigmaoned\ values, shown in Figure \ref{cl_lw}, are equivalent to the clump internal velocity dispersions and shed light on the turbulent nature of the clump interiors. 

We compare the \sigmaoned\ values for starless and protostellar clumps in Table \ref{intdisp}, to determine if there is a statistically significant difference between the two types of objects. We define a particular M01 core as being protostellar if it has an IRAC Class I match in NWT, or if M01 have designated them as protostellar. Our \emph{clumps} are then deemed to be protostellar if a protostellar M01 core lies within the extent of the ellipse defined by the \gclumps\ algorithm. We also narrow our range of starless clumps down to only include those that are associated with an M01 core, all others are deemed to be `noise' or transient clumps. Using these definitions, we find 4 protostellar and 3 starless \HthirteenCOplus\ clumps; and 5 protostellar and 10 starless \CeighteenO\ clumps. We find that in the case of \HthirteenCOplus, the \sigmaoned\ values are much lower for starless clumps than for protostellar clumps; the \CeighteenO\ \sigmaoned\ values also show the same trend, but to a lesser extent.

We performed a Kolmogorov-Smirnoff (KS) test, to determine if the two protostellar and starless datasets were drawn from the same underlying distribution. The KS statistic (or p-value) gives a value that is 1 minus the confidence level with which the null hypothesis (that the two samples originate from the same distribution) may be rejected. Generally, if the p-value is $<$ 1\%, we can say confidently that the two samples originate from different underlying distributions. KS tests give p-values of 15\% and 26\% for \HthirteenCOplus\ and \CeighteenO\ respectively, when comparing protostellar and starless clump distributions. Therefore at the 85\% confidence level, the \HthirteenCOplus\ protostellar and starless clumps originate from different underlying distributions, justifying our clump classification system. There is a greater difference between the two distributions in the case of \HthirteenCOplus\ as it traces the denser gas in the cores, which should be more quiescent and also contain less centrally condensed gas for starless cores than protostellar cores; the \CeighteenO\ on the other hand contains more of a contribution from the core envelopes (as mentioned previously) making the distributions less distinct.

\begin{table*}
\centering
\caption{Mean internal 1D velocity dispersions of the Protostellar (P) and Starless (S) \CeighteenO\ and \HthirteenCOplus\ clumps; as well as for the two molecules without differentiating clump type. The errors on the last digit (standard deviations on the sample) are given in brackets after each value.}
\begin{tabular}{ccccccc}
\hline
Dispersion & \HthirteenCOplus\ (P) & \CeighteenO\ (P) & \HthirteenCOplus\ (S) & \CeighteenO\ (S) & \HthirteenCOplus\ (All) & \CeighteenO\ (All)\\ \hline
\sigmaoned/\kms\ & 0.30(8) & 0.35(6) & 0.17(5) & 0.30(5) & 0.27(10) & 0.28(8)\\
\sigmaoned/$c_s$ & 1.4(4) & 1.47(23) & 0.86(22) & 1.25(22) & 1.10(4) & 1.2(3)\\
\sigmaoned/$\sigma_g$ & 0.55(17) & 0.58(10) & 0.34(9) & 0.49(9) & 0.45(17) & 0.46(2)\\
\hline
\end{tabular}
\label{intdisp}
\end{table*}

The \sigmaoned\ values are also compared to the sound speed $c_{\rm s}$ in the region, to determine how turbulent the clump interiors are. We define three turbulent regimes: subsonic (\sigmaoned\ $<$ $c_{\rm s}$), transonic ($c_{\rm s} <$ \sigmaoned\ $< 2c_{\rm s}$) and supersonic ($2c_{\rm s} <$ \sigmaoned). The value of $c_{\rm s}$ is calculated using the following formula:
\begin{equation}
c_{\rm s} = \sqrt{\frac{kT_{\rm ex}}{m}}
\label{cs}
\end{equation}
$T_{\rm ex} = 16.1 $ K is the value adopted earlier in Section 5.4 , and $m = 2.33 m_{\rm p}$, assuming a ratio of 1 Helium for every 5 H$_2$ molecules, giving a value of $c_{\rm s} = 0.24$ \kms. We also calculate the bulk gas velocity dispersion $\sigma_g = 0.61 \pm 0.29$ \kms\ (from the \thirteenCO\ gas over the region), to compare the clumps' internal turbulence with their surrounding natal gas environment.

We find that the \sigmaoned\ values for all clumps are transonic, except for the starless \HthirteenCOplus\ clumps which are subsonic. This is consistent with the starless clumps corresponding to local minima of turbulence, as they are much less turbulent internally than the surrounding gas. In addition, we see that the \sigmaoned\ values for all clumps are much smaller than the bulk gas velocity dispersion. These results are consistent with the predictions of gravoturbulent fragmentation --- {i.e.} that the protostellar cores are found at the intersection points of turbulent shocks and are therefore much less turbulent internally than their surrounding natal gas. \changed{However, we do note that most star formation scenarios agree that the non-thermal support should decrease in star-forming cores, so this is not conclusive observational evidence of gravoturbulent fragmentation.}

\subsubsection{Relative core-gas motions}
The interclump velocity dispersion \sigmactc\ that we calculate is the dispersion (or standard deviation) of the line centre velocities v$_{\rm c}$ at the clump centres, for all the clumps in the region. Initially, without distinguishing between protostellar and starless clumps, we find that the \ctcdisp\ decrease as the density of gas traced by the molecule increases (see Figure \ref{vcom}): $0.39 \pm 0.05$ \kms\ and $0.28 \pm 0.08$ \kms\ for \CeighteenO\ and \HthirteenCOplus\ respectively. A KS test comparing \sigmactc\ for the two molecules indicates with 85\% confidence that they are tracing gas from different underlying distributions. \citet{2007A&A...472..519A} calculated a core-to-core dispersion of $\sim$ 0.25 \kms\ for protostellar cores in the L1688 region of Ophiuchus \changed{using N$_2$H$^+$, which traces similar densities of gas} and is very close to the value we calculate for \HthirteenCOplus\ clumps, but only about two-thirds of that for the \CeighteenO\ clumps. This discrepancy could be explained however by the difference in global properties of the overall molecular cloud. Orion is a much more turbulent, clustered star-forming environment than Ophiuchus and the differences between bulk gas motion and individual star-forming cores might be more pronounced. The \CeighteenO\ clumps cover much larger areas and would be more likely to be influenced by the bulk gas movements, whereas the \HthirteenCOplus\ clumps are tracing denser regions and are more likely to reflect the dispersions between star-forming cores.

\begin{table*}
\caption{Interclump velocity dispersions of the Protostellar (P) and Starless (S) \CeighteenO\ and \HthirteenCOplus\ clumps; as well as for the two molecules without differentiating clump type. The errors on the last digit(s) are given in brackets after each value, and are calculated using the formula from \citet{2007A&A...472..519A}.}
\centering
\begin{tabular}{ccccccc}
\hline
Dispersion & \HthirteenCOplus\ (P) & \CeighteenO\ (P) & \HthirteenCOplus\ (S) & \CeighteenO\ (S) & \HthirteenCOplus\ (All) & \CeighteenO\ (All) \\ \hline
\sigmactc/\kms\ & 0.27(11) & 0.35(12) & 0.22(11) & 0.49(11) & 0.28(8) & 0.39(5)\\
\sigmactc/$c_s$ & 1.14(3) & 1.45(3) & 0.92(3) & 2.03(3) & 1.17(2) & 1.61(2)\\
\sigmactc/$\sigma_g$ & 0.45(14) & 0.57(18) & 0.36(11) & 0.80(25) & 0.46(14) & 0.63(19)\\
\hline
\end{tabular}
\label{ctcdisp}
\end{table*}

Using the same selection criteria as before, we compare the \ctcdisp\ between protostellar and starless clumps.  The \ctcdisp\ were also compared with the region sound speed and the region bulk-gas velocity, in the same manner as in the previous section. The results are presented in Table \ref{ctcdisp}. 

All clumps show transonic behaviour (except the starless \HthirteenCOplus\ clumps which are subsonic), indicating that there is little discernable movement between clumps. Comparing the \ctcdisp\ with the bulk-gas velocity dispersion, we find that the motion of the clumps compared to their surrounding gas is fairly low, with \sigmactc/$\sigma_g$ values of 0.46 and 0.80 for \HthirteenCOplus\ and \CeighteenO\ respectively. This means that the clumps are in some way still coupled to the surrounding gas from which they formed, and indicates that the clumps should be able to move together with their surrounding gas under the same gravitational forces, one of the key requirements for competitive accretion to occur \citep{2006MNRAS.370..488B}. The relative clump-gas velocity for \HthirteenCOplus\ is lower than that for \CeighteenO, which is most likely due to the \CeighteenO\ tracing more of the surrounding envelope that is expected to be more turbulent, and hence will retain a greater degree of the movement of the natal gas.

\changed{\citet{2008AJ....136..404O} compared core-to-core velocity dispersions for observational data taken in the Perseus \citep{2007ApJ...668.1042K} and $\rho$ Ophiuchus regions \citep{2007A&A...472..519A}, with those produced in simulations of driven and decaying turbulence. They found that the driven run was in better agreement with Perseus while the decaying run gave a better match for Ophiuchus. They did however note that the two simulations were statistically very similar, and that in general the distribution of core central velocities did not appear to depend on the details of the turbulence. When they differentiated between prestellar and protostellar cores in their simulations, they were able to distinguish between the two turbulent scenarios.} They found that in driven turbulence simulations, the \emph{prestellar} core-to-core velocity dispersions were larger than those for protostellar cores; conversely for decaying turbulence simulations, the \emph{protostellar} core-to-core velocity dispersions are larger than those for the prestellar cores. In both cases, the difference is about a factor of 1.3. 

Comparing \ctcdisp\ for our protostellar and starless clumps (from the same molecule), we find that \HthirteenCOplus\ protostellar clumps have a higher \sigmactc\ than the starless clumps, indicative of decaying turbulence; the opposite is true for \CeighteenO, where \sigmactc\ is higher for starless clumps than protostellar clumps. From KS tests, we cannot rule out the possibility that the protostellar and starless clumps originate from the same underlying distribution (p-values of 26\% and 54\% for \HthirteenCOplus\ and \CeighteenO\ respectively). The \CeighteenO\ gas may also be more affected by other factors such as multiple velocity components along the line-of-sight or outflows, which could all potentially confuse the line centre velocity. We therefore believe that the \HthirteenCOplus\ will give a better picture of the star-forming cores themselves, given that they are tracing the denser gas. \changed{In comparison, core-to-core velocity dispersions of N$_2$H$^+$ observations in Perseus \citep{2007ApJ...668.1042K}, provide support for driven turbulence --- they calculated a higher \sigmactc\ for their prestellar cores than for their protostellar cores. As their emission traces similar densities to the \HthirteenCOplus\ in this analysis, the type of turbulence present in Perseus is therefore different from that present in NGC~2068.} However, as we only have a small sample of \HthirteenCOplus\ clumps, we cannot say for certain whether decaying turbulence is dominant in our region; a larger statistical sample (preferably over different regions, to confirm environmental effects) would be required to confirm the results.


\subsubsection{Comparison with Decaying Turbulence Numerical Simulation}
The main numerical simulation that we are comparing our data with, is that of SCB09. The simulation consists of a 3-dimensional cylinder (3 $\times$ 3 $\times$ 10 pc) containing 10$^4$ \Msun\ concentrated at one end, so the top is over-bound and the bottom is under-bound. The turbulence in this simulation is not driven but allowed to decay once self-gravity is applied; however it is replenished by the release of kinetic energy as the molecular cloud collapses. Further details of the simulation can be found in the SCB09 paper. A catalogue of p-cores (found via a gravitational potential clumpfinding algorithm, as mentioned in Section 4.3) is produced at snapshot intervals of 0.1 $t_{\rm dyn}$ (4.7 $\times$ 10$^4$ years). For each p-core, the catalogue contains information about the centre-of-mass (CoM) velocities and positions, and their masses, as well as their potential, kinetic and thermal energies. In their simulation, SCB09 have defined a subset of `bound' p-cores, for which $E_{\rm rat} >$ 1, where $E_{\rm rat}$ is defined as:

\begin{equation}
E_{\rm rat} = \frac{\| E_{\rm p} \|}{E_{\rm therm} + E_{\rm K}},
\end{equation}
and $E_{\rm therm}$ is the thermal energy of the clump, $E_{\rm K}$ is the kinetic energy calculated with respect to the CoM velocity of the clump and $E_{\rm p}$ is the potential energy of the clump calculated from the depth of the background-subtracted potential well. These p-cores are therefore bound with respect to the environment in which they were formed, and not merely in isolation (as is the case for the clumps we identified using traditional clumpfinding methods). Bound p-cores are assumed to be pre-stellar in this simulation, as they go on to form sinks that accumulate mass and serve as the observable manifestation of star formation. As such, we would like to make velocity dispersion comparisons of our \HthirteenCOplus\ and \CeighteenO\ clumps with these pre-stellar cores. 

We use p-core catalogues from 4 snapshots in time: 1.025, 1.050, 1.125 and 1.180 $t_{\rm dyn}$, corresponding to a span of time between 4.7 $\times$ 10$^5$ and 5.6 $\times$ 10$^5$ years. We combined the 4 catalogues into a single data-set, to eliminate any time-dependent effects and to increase the sample size. A KS test performed on the 4 snapshots gave p-values $>$ 20 \%, implying that they originate from different underlying distributions, and we are justified in assuming the samples to be independent when combining them. We also refine our sample by only choosing bound p-cores, using the same criterion as SCB09, that E$_{\rm rat} >$ 1. Our final dataset consists of 533 bound prestellar p-cores, which is large enough to be statistically meaningful.

Initially, we calculate the intercore velocity dispersion \sigmactc\ of all the p-cores in the simulation, in the same way as we did before. We find that \sigmactc\ = $0.602 \pm 0.017$ \kms, which is much larger than the individual \ctcdisp\ of the \HthirteenCOplus\ and \CeighteenO\ clumps. 

\begin{figure}
\centerline{\includegraphics[width=3.9cm,height=3.9cm]{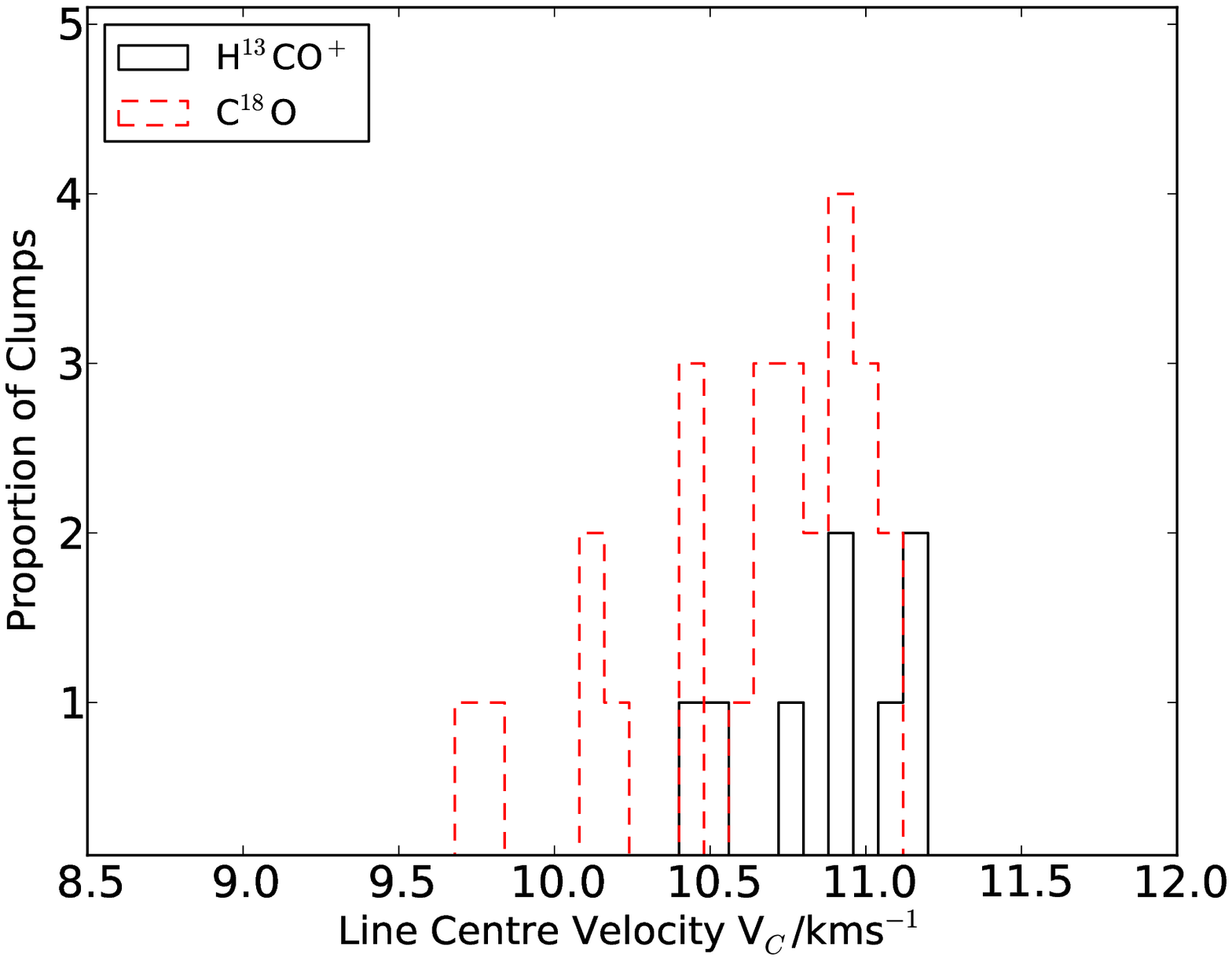}\qquad
	    \includegraphics[width=4.1cm,height=3.9cm]{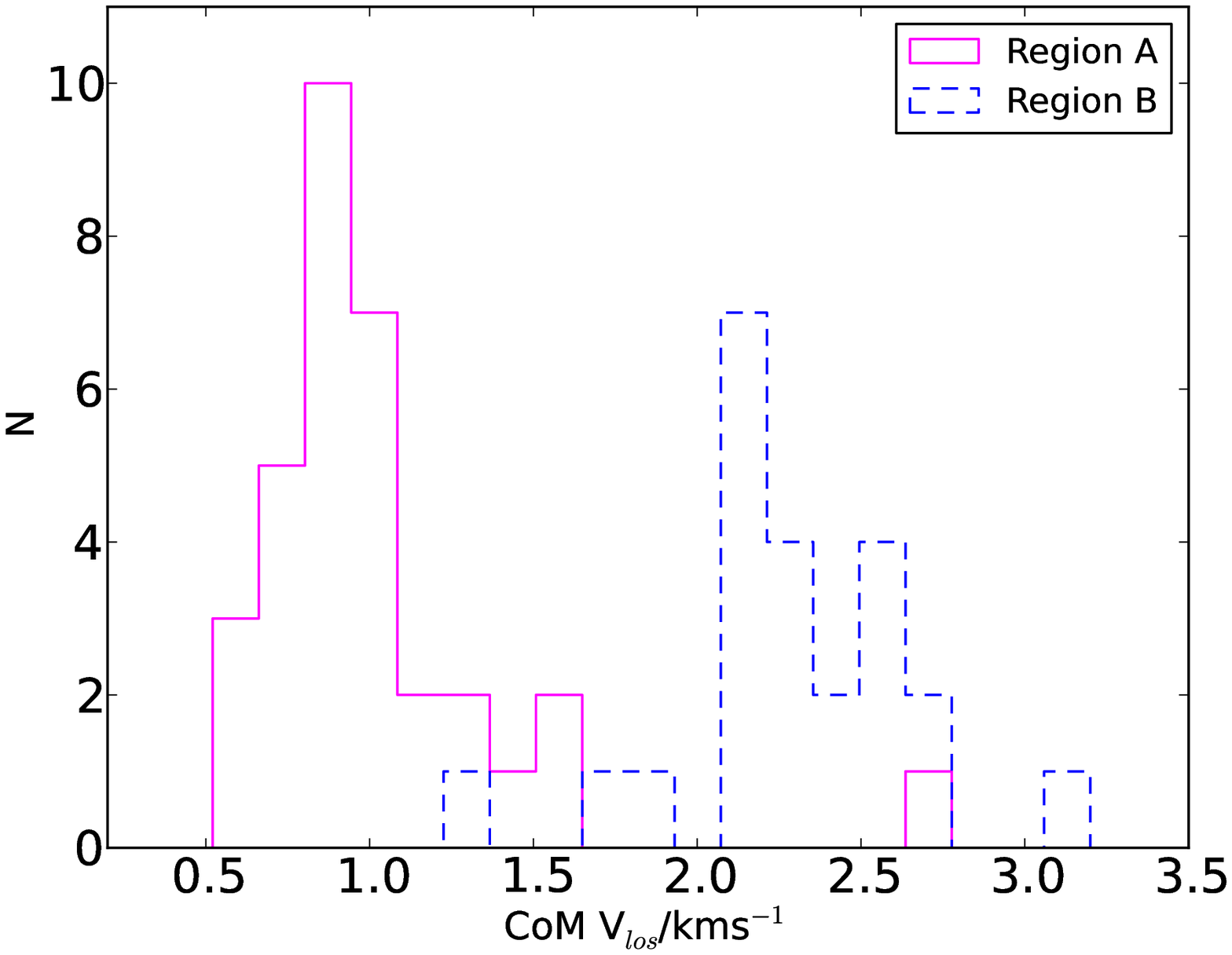}}
\caption{\changed{Histograms showing the spread of clump line centre velocities for the \HthirteenCOplus\ and \CeighteenO\ clumps (left); and the equivalent spread of line-of-sight velocities calculated for the bound p-cores in SCB09 (right).}}
\label{vcom}
\end{figure}

To better match the size of NGC~2068, we then extract 2 parsec-sized regions, which are taken from different parts of the simulation, and so have differently-bound initial environments. Region A correponds to the initially less-bound environment, while Region B corresponds to the initially more-bound environment. Figure \ref{vcom} shows a comparison between the observed and simulated intercore velocity dispersions. We calculate \sigmactc\ for the p-cores within each region, and obtain values of $0.41 \pm 0.05$ \kms\ and $0.36 \pm 0.05$ \kms\ for Regions A and B respectively. The values calculated for the smaller regions are very close to the \ctcdisp\ for the \CeighteenO\ clumps (see Table \ref{ctcdisp}), which makes sense as the size scales over which we are calculating the dispersions are similar. The \HthirteenCOplus\ clumps are concentrated in a smaller area than the \CeighteenO\ clumps, and this is reflected in their \ctcdisp, which are 70\% of the p-core \sigmactc\ values. We therefore suggest that when calculating and comparing \ctcdisp, the size of the region considered is an important factor, as the \sigmactc\ appears to vary with region size. \changed{This is similar to the Larson relation \citep{1981MNRAS.194..809L} of $\sigma$(\kms)$=1.10 L(pc)^{0.38}$, which reflects the underlying turbulence scaling between \sigmactc\ and region size.}

We find that the simulation --- which models a decaying turbulence environment --- is able to reproduce quite closely the effects and properties that are seen observationally, when comparing similar size-scales. However, there is still insufficient evidence to say conclusively that the region we observe is dominated by dynamically decaying turbulence.

%

\section{Conclusions}\label{sec:concl}
We have presented \thirteenCO, \CeighteenO\ and \HthirteenCOplus\ data for the NGC~2068 region in Orion B, the combination of which allows us to probe different densities and size-scales, and determine the structural and kinematic properties of the region. 

The \thirteenCO\ is widespread and detected over the entire region, while \CeighteenO\ has a more filamentary structure and follows the SCUBA dust emission much more closely; the \HthirteenCOplus\ is confined to small clumps, the majority of which are found in the region centre. The \thirteenCO\ and \CeighteenO\ show evidence of multiple velocity components along the line-of-sight, especially in the west of the region; the \HthirteenCOplus\ remains single peaked throughout, with a line centre velocity of $10.86 \pm 0.10$ \kms, reflecting the systemic velocity of the region. The \thirteenCO\ also shows evidence of several bipolar outflows in the region, and we have matched these to SCUBA cores, as far as possible given the density of sources in the central region.

We have used the \gclumps\ algorithm to decompose the \CeighteenO\ and \HthirteenCOplus\ data into Gaussian clumps, and calculated various clump properties, including size/shape, linewidths, and virial and gas masses. The \CeighteenO\ clumps have higher linewidths than the \HthirteenCOplus\ clumps on average, and are significantly larger than the \HthirteenCOplus\ clumps. We therefore conclude that the \HthirteenCOplus\ traces the denser inner parts of the star-forming cores, while the \CeighteenO\ likely traces more of the less dense, outer envelopes of the cores. Using the ratio of virial to gas mass, we have found that the \CeighteenO\ clumps mostly appear more bound than the \HthirteenCOplus\ clumps. However this may not be the best method of calculating how bound a particular clump is --- the \CeighteenO\ clumps appear more bound as there is a greater contribution from the surrounding envelope; using the dust mass instead of the gas mass gives an alternative view of the mass contained within the \HthirteenCOplus\ clumps, making them appear much more bound.

We have classified our molecular clumps as protostellar or starless based on a match with a known YSO or SCUBA core respectively. We have also used the internal and interclump velocity dispersions for the two molecules to investigate star formation mechanisms. The internal velocity dispersions of the starless \HthirteenCOplus\ clumps (probably the most likely objects to represent prestellar cores) are subsonic, suggesting that the cores were formed at local turbulence minima, consistent with gravoturbulent fragmentation. The low values of the interclump velocity dispersions compared to the bulk gas dispersion (as traced by the \thirteenCO) also indicate a certain degree of coupling between the possible star-forming cores and their natal gas environment, fulfilling one of the key requirements for competitive accretion to occur. The interclump velocity dispersions match numerical simulations of decaying turbulence, when similar sized regions are compared; the \HthirteenCOplus\ protostellar and starless interclump velocity dispersions also match predictions for decaying turbulence well. However, due to the small size of our sample, we cannot say for certain that decaying turbulence is dominant in the region. A larger statistical sample (over regions with different gas environments) would be required to conclusively determine the type and decay timescale of the turbulence present in this region, and how the nature of the turbulence varies between regions.

\section*{Acknowledgments}
S. Walker-Smith is funded by the Science and Technology Facilities Council of the UK. The JCMT is supported by the Science and Technology Facilities Council of the UK, and the National Research Council Canada, and the Netherlands Organisation for Scientific Research.


\appendix
\section{Clump catalogues}

A catalogue of the clumps found by \gclumps\ for the \HthirteenCOplus\ and \CeighteenO\ data is provided in this appendix, in Tables \ref{tab:hcoprop} and \ref{tab:c18oprop} respectively.

\begin{table*}
\caption{Catalogue of \gclumps\ clumps found for \HthirteenCOplus\ emission in NGC~2068. Column 2: Deconvolved FWHM size of each clump in the 2 spatial dimensions; Column 3: Peak intensity in each clump; Column 4: Internal velocity dispersion \sigmaoned\ at the peak of each clump; Column 5: Line centre velocity at the peak of each clump; Column 6: Ratio of the FWHM sizes of each clump; Column 7: Effective radius of the clump; Column 8: Virial mass of the clump; Column 9: The sum of values extracted from within the ellipse defined by the clump FWHM; Column 10: Mass of clump calculated using the values in column 9 assuming LTE; Column 11: Ratio of the gas mass to the virial mass; Column 12: Mass of the M01 SCUBA cores associated with each clump.}
\begin{center}
\begin{tabular}{cccccccccccc}
\hline
Clump & Size & Peak & \sigmaoned\ & $v_{\rm c}$ &  Axis & $r_{\rm eff}$ & $M_{\rm vir}$ & Clump & $M_{\rm gas}$ & Bound & $M_{\rm smm}$\\
No. & /(AU $\times$ AU) & /K & /\kms\ & /\kms\ & ratio & /pc & /\Msun\ & sum /K & /\Msun\ & ratio & /\Msun\ \\
\hline
1 & 2200 $\times$ 2100 & 0.55 & 0.28 & 10.54 & 0.99 & 0.022 & 0.73 & 53.6 & 0.93 & 1.3 & 3.00\\
2 & 2600 $\times$ 8200 & 0.50 & 0.19 & 11.05 & 0.63 & 0.037 & 0.42 & 63.8 & 1.10 & 2.6 & 1.05\\
3 & 11000 $\times$ 3300 & 0.42 & 0.22 & 10.94 & 0.55 & 0.044 & 0.89 & 48.1 & 0.83 & 0.9 & 3.50\\
4 & 3500 $\times$ 3100 & 0.42 & 0.43 & 10.93 & 0.97 & 0.026 & 1.63 & 53.5 & 0.92 & 0.6 & 1.35\\
5 & 8900 $\times$ 5500 & 0.35 & 0.16 & 11.17 & 0.75 & 0.046 & 0.34 & 80.5 & 1.39 & 4.1 & 0.40\\
6 & 5000 $\times$ 2400 & 0.26 & 0.42 & 10.40 & 0.82 & 0.029 & 2.04 & 51.0 & 0.88 & 0.4 & 1.35\\
7 & 5200 $\times$ 3200 & 0.22 & 0.21 & 11.14 & 0.85 & 0.031 & 0.80 & 40.3 & 0.70 & 0.9 & ---\\
8 & 7400 $\times$ 5400 & 0.22 & 0.27 & 10.74 & 0.84 & 0.042 & 0.93 & 33.4 & 0.58 & 0.6 & 0.90\\
\hline
\end{tabular}
\end{center}
\label{tab:hcoprop}
\end{table*}

\begin{table*}
\caption{Catalogue of \gclumps\ clumps found for \CeighteenO\ emission in NGC~2068. Column 2: Deconvolved FWHM size of each clump in the 2 spatial dimensions; Column 3: Peak intensity in each clump; Column 4: Internal velocity dispersion \sigmaoned\ at the peak of each clump; Column 5: Line centre velocity at the peak of each clump; Column 6: Ratio of the FWHM sizes of each clump; Column 7: Effective radius of the clump; Column 8: Virial mass of the clump; Column 9: The sum of values extracted from within the ellipse defined by the clump FWHM; Column 10: Mass of clump calculated using the values in column 9 assuming LTE; Column 11: Ratio of the gas mass to the virial mass; Column 12: Mass of the M01 SCUBA cores associated with each clump.}
\begin{center}
\begin{tabular}{cccccccccccc}
\hline
Clump & Size & Peak & \sigmaoned\ & $v_{\rm c}$ &  Axis & $r_{\rm eff}$ & $M_{\rm vir}$ & Clump & $M_{\rm gas}$ & Bound & $M_{\rm smm}$\\
No. & /(AU $\times$ AU) & /K & /\kms\ & /\kms\ & ratio & /pc & /\Msun\ & sum /K & /\Msun\ & ratio & /\Msun\ \\
\hline
1 & 15000 $\times$ 13000 & 10.68 & 0.34 & 11.04 & 0.88 & 0.083 & 1.29 & 3424 & 11.89 & 9.2 & 2.45 \\
2 & 14000 $\times$ 16000 & 8.13 & 0.36 & 10.88 & 0.88 & 0.088 & 1.27 & 2960 & 10.28 & 8.1 & 2.00 \\
3 & 13000 $\times$ 15000 & 7.06 & 0.20 & 10.90 & 0.87 & 0.082 & 0.60 & 1793 & 6.22 & 10.4 & --- \\
4 & 7800 $\times$ 4000 & 6.59 & 0.22 & 10.71 & 0.74 & 0.038 & 0.33 & 546 & 1.90 & 5.8 & --- \\
5 & 17000 $\times$ 13000 & 6.14 & 0.42 & 10.75 & 0.78 & 0.087 & 2.72 & 2645 & 9.18 & 3.4 & 3.55 \\
6 & 17000 $\times$ 12000 & 5.78 & 0.30 & 10.68 & 0.74 & 0.084 & 2.36 & 1934 & 6.71 & 2.8 & 0.60 \\
7 & 7900 $\times$ 10000 & 4.95 & 0.34 & 10.46 & 0.82 & 0.055 & 1.96 & 1081 & 3.75 & 1.9 & 0.65 \\
8 & 12000 $\times$ 9800 & 5.19 & 0.20 & 11.01 & 0.84 & 0.065 & 0.57 & 388 & 1.35 & 2.4 & --- \\
9 & 6200 $\times$ 7700 & 5.05 & 0.40 & 10.17 & 0.88 & 0.043 & 0.67 & 930 & 3.23 & 4.8 & 1.30 \\
10 & 8000 $\times$ 11000 & 4.70 & 0.21 & 10.65 & 0.78 & 0.058 & 0.75 & 258 & 0.89 & 1.2 & --- \\
11 & 11000 $\times$ 9900 & 4.80 & 0.20 & 10.42 & 0.93 & 0.062 & 0.40 & 570 & 1.98 & 4.9 & --- \\
12 & 7200 $\times$ 9700 & 4.61 & 0.25 & 9.79 & 0.83 & 0.051 & 0.66 & 995 & 3.45 & 5.3 & 1.90 \\
13 & 11000 $\times$ 2900 & 5.77 & 0.27 & 10.62 & 0.53 & 0.043 & 0.61 & 577 & 2.00 & 3.3 & --- \\
14 & 21000 $\times$ 14000 & 4.31 & 0.25 & 11.02 & 0.72 & 0.101 & 1.02 & 2001 & 6.95 & 6.8 & 1.20 \\
15 & 9600 $\times$ 8900 & 7.28 & 0.31 & 11.01 & 0.95 & 0.056 & 0.47 & 1257 & 4.36 & 9.3 & 2.00 \\
16 & 11000 $\times$ 7900 & 3.98 & 0.19 & 10.91 & 0.77 & 0.057 & 0.46 & 583 & 2.02 & 4.4 & --- \\
17 & 7600 $\times$ 5700 & 3.72 & 0.23 & 9.69 & 0.85 & 0.042 & 0.67 & 448 & 1.56 & 2.3 & 4.30 \\
18 & 30000 $\times$ 5100 & 2.82 & 0.27 & 10.81 & 0.26 & 0.083 & 0.99 & 575 & 2.00 & 2.0 & 0.60 \\
19 & 7300 $\times$ 13000 & 2.86 & 0.16 & 10.76 & 0.66 & 0.059 & 0.54 & 577 & 2.00 & 3.7 & --- \\
20 & 7800 $\times$ 8100 & 3.45 & 0.42 & 10.10 & 0.97 & 0.049 & 1.26 & 822 & 2.85 & 2.3 & --- \\
21 & 3400 $\times$ 7000 & 2.69 & 0.28 & 10.12 & 0.75 & 0.035 & 0.26 & 392 & 1.36 & 5.3 & 2.50 \\
22 & 9500 $\times$ 17000 & 2.44 & 0.32 & 10.91 & 0.63 & 0.076 & 1.33 & 1163 & 4.04 & 3.1 & 2.30 \\
23 & 3900 $\times$ 4400 & 2.52 & 0.24 & 10.78 & 0.97 & 0.03 & 0.34 & 187 & 0.65 & 1.9 & --- \\
24 & 15000 $\times$ 4200 & 2.52 & 0.23 & 11.11 & 0.45 & 0.055 & 0.30 & 486 & 1.69 & 5.7 & --- \\
25 & 7100 $\times$ 4600 & 4.23 & 0.39 & 10.84 & 0.82 & 0.038 & 0.17 & 629 & 2.18 & 13.1 & 1.35 \\
26 & 7300 $\times$ 3200 & 5.75 & 0.30 & 10.45 & 0.72 & 0.035 & 0.39 & 531 & 1.84 & 4.8 & 0.90 \\
\hline
\end{tabular}
\end{center}
\label{tab:c18oprop}
\end{table*}

\label{lastpage}

\end{document}